\documentclass[twocolumn,prb,asmath,floatfix]{revtex4}
\usepackage{lscape}
\usepackage{epsf}
\usepackage{ulem}
\usepackage[dvips]{graphicx}
\usepackage{rotating}
\usepackage{bm}
\usepackage{color}
\usepackage{amsmath}
\usepackage{cancel}
\usepackage{ulem}
\usepackage{color}
\definecolor{Blue}{rgb}{0,0,1}
\definecolor{Red}{rgb}{1,0,0}
\definecolor{Green}{rgb}{0,0.52,0.0}
\definecolor{orange}{rgb}{1,0.5,0}
\definecolor{orange2}{rgb}{1,0.5,0.5}

\date{\today}

\usepackage{amssymb}

\providecommand{\abs}[1]{\lvert #1 \rvert}

\DeclareMathOperator{\sgn}{sgn}
\newcommand{\lbl}{\label}
\newcommand{\Hamiltonian}{\mathcal{H}}  

\begin{document}

\title{Dzyaloshinskii-Moriya interaction and chiral magnetism in 3$d$-5$d$ zig-zag chains: Tight-binding model and \textit{ab initio} calculations}

\author{Vikas Kashid}
\affiliation{Department of Physics, University of Pune, Pune 411007, India}
\author{Timo Schena}
\affiliation{Peter Gr\"{u}nberg Institut and Institute for Advanced Simulation, Forschungszentrum J\"{u}lich and JARA, D-52425 J\"{u}lich, Germany}
\author{Bernd Zimmermann}
\affiliation{Peter Gr\"{u}nberg Institut and Institute for Advanced Simulation, Forschungszentrum J\"{u}lich and JARA, D-52425 J\"{u}lich, Germany}
\author{Vaishali Shah}
\affiliation{Interdisciplinary School of Scientific Computing, University of Pune, Pune 411007, India} 
\author{H. G. Salunke}
\affiliation{Technical Physics Division, Bhabha Atomic Research Centre, Mumbai 400085 India}
\author{Yuriy Mokrousov}
\email{y.mokrousov@fz-juelich.de}
\affiliation{Peter Gr\"{u}nberg Institut and Institute for Advanced Simulation, Forschungszentrum J\"{u}lich and JARA, D-52425 J\"{u}lich, Germany}
\author{Stefan Bl\"{u}gel}
\email{s.bluegel@fz-juelich.de}
\affiliation{Peter Gr\"{u}nberg Institut and Institute for Advanced Simulation, Forschungszentrum J\"{u}lich and JARA, D-52425 J\"{u}lich, Germany}

\begin{abstract}
We investigate the chiral magnetic order in free-standing planar 3$d$-5$d$ bi-atomic metallic chains (3$d$: Fe, Co; 5$d$: Ir, Pt, Au) using first-principles calculations based on density functional theory.  We found that the antisymmetric exchange interaction, commonly known as Dzyaloshinskii-Moriya interaction (DMI), contributes significantly to the energetics of the magnetic structure.  For the Fe-Pt and Co-Pt chains, the DMI can compete with the isotropic Heisenberg-type exchange interaction and the magneto-crystalline anisotropy energy (MAE), and for both cases a homogeneous left-rotating cycloidal chiral spin-spiral with a wave length of $51$ \AA\ and $36$ \AA, respectively, were found. The sign of the DMI, that determines the  handedness of the magnetic structure changes in the sequence of the $5d$ atoms Ir($+$), Pt($-$), Au($+$).  We used the full-potential linearized augmented plane wave method and performed self-consistent calculations of homogeneous spin spirals, calculating the DMI by treating the effect of spin-orbit interaction (SOI) in the basis of the spin-spiral states in first-order perturbation theory.
To gain insight into the DMI results of our \textit{ab initio} calculations, we develop a minimal tight-binding model of three atoms and 4 orbitals that contains all essential features: the  spin-canting between the magnetic $3d$ atoms, the spin-orbit interaction at the $5d$ atoms, and the  structure  inversion asymmetry facilitated by the  triangular geometry. We found that spin-canting can lead to spin-orbit active eigenstates that split in energy due to the spin-orbit interaction at the $5d$ atom. We show that, the sign and strength of the  hybridization, the bonding or antibonding character between $d$-orbitals of the magnetic and non-magnetic sites, the bandwidth and the energy difference between states occupied and unoccupied states of different spin projection  determine the sign and strength of the DMI. The key features observed in the trimer model are also found in the first-principles results.

\end{abstract}

\maketitle
\section{Introduction}

The recent discovery of chiral magnetism in low-dimensional metals\cite{Bode} has opened a new vista in the research of magnetism. For a two-dimensional Mn monolayer film on W(110) it was shown that the magnetic structure was not the two-dimensional checkerboard antiferromagnetic one\cite{Heinze:00} as thought of for a long time, instead by combining spin-polarized scanning tunneling microscopy and {\it ab initio} theory a left-rotating cycloidal spin spiral was found. A right-rotating one, which would have the same energy in a conventional achiral magnet, does not exist. Since then, chiral magnetism was found in several other thin-film system Mn/W(100)\cite{ferriani},  and in bi-atomic Fe chains on the ($5\times1$)-Ir(001) surface\cite{Menzel_12}. Chiral magnetism was recently also found in domain walls, {\it e.g.}\ in Fe/W(110),\cite{Kubetzka_02, heidem} Ni/Fe/Cu(001),\cite{YZWu_13} Co/Pt(111),\cite{Freimuth_14} Co/Pt,\cite{Parkin_13} FeCo/Pt\cite{Beach_13} and in the magnon dispersion of Fe/W(110)\cite{Kirschner_10}.  In most cases the chirality is imprinted in one-dimensional chiral spin spirals, but under certain conditions chirality can also appear in form of two-dimensional objects known as skyrmions, {\it e.g.}\  in case of Fe/Ir(111)\cite{Bergmann_06, Bergmann_07, Heinze_11} and  Pd/Fe/Ir(111).\cite{Science_13} The chirality in these low-dimensional magnets opens completely new perspectives in domain-wall motion, spin-torques or  spin transport that all together have really an impact on the further development of spintronics. 

The origin of the chirality in low-dimensional  itinerant magnets is caused by the presence of the spin-orbit interaction (SOI) in combination with a  
structure inversion asymmetry provided by a substrate on which the film is deposited. This leads to an antisymmetric exchange interaction, postulated first by Dzyaloshinskii\cite{Dzyaloshinsky:58.1} and frequently referred to as the Dzyaloshinskii-Moriya-type interaction (DMI), because Moriya\cite{moriya}
provided the first microscopic understanding on the basis of a model relevant to insulators.  Although the microscopic models for metals are naturally different and go back to Smith\cite{Smith:76.1}, Fert and Levy\cite{Fert:80.1,Fert:90.1} and  Kataoka {\it et al.}\cite{Kataoka:84.1},
the functional form of the antisymmetric exchange remains unchanged. If the Dzyaloshinskii-Moriya-type interaction is sufficiently strong, it can compete with the conventional isotropic exchange interaction of spins and the magneto-crystalline anisotropy,
and the conventional ferromagnetic or antiferromagnetic phase is destabilized in favor of a chiral one. The isotropic exchange interaction goes back to the Coulomb interaction in combination with the antisymmetric nature of the many-electron wave function and the hopping of  electrons. It is  typically captured by the Heisenberg model. The Heisenberg  interaction is strictly achiral and any spiral state produced by the Heisenberg interaction is symmetric with respect to left- or right chirality.  Whether the Dzyaloshinskii-Moriya-type interaction is strong enough to stabilize a chiral spiral and which sign the interaction will take on, determining the chirality of the rotating structure (right- or left-rotating),  is {\it a priori} unknown and depends on the details of the electronic structure.
 
Homogeneous and inhomogeneous chiral spirals have been investigated by Dzyaloshinskii\cite{Dzyaloshinsky:65:2} on a model level. Surprisingly, little is known quantitatively about the Dzyaloshinskii-Moriya-type interaction in low-dimensional metallic magnets. Practically no systematic theoretical or computational results exist. Obviously, it is a chiral interaction based on spin-orbit interaction and requires the treatment of non-collinear magnetism in a broken symmetry environment, which necessitates typically the computation of small quantities in a complex geometry. In particular, this interaction is small as compared to the Heisenberg exchange, and therefore we expect chiral spirals of long wave lengths that deviate little from the ferromagnetic state. Thus, in terms of {\it ab initio} calculations  it means an accurate  treatment requires precise calculations of gigantic unit cells that are unattainable even with modern supercomputers.  All-in-all, this makes the treatment rather non-trivial. 

In this paper we shed light onto the DMI by performing calculations based on the density functional theory to a  well-chosen set of model systems, namely  planar  free-standing  zigzag bi-atomic chains of 3$d$ and 5$d$ transition-metal atoms in a structure inversion asymmetric geometry.  That means, we have chosen  a combination of  $3d$ elements (Fe or Co) exhibiting strong magnetism and heavy $5d$ elements (Ir, Pt or Au) as source of strong SOI.
 The asymmetric chain can be considered as a minimal model describing a film of 3$d$ atoms
on a non-magnetic substrate with large spin-orbit interaction, or a chain of $3d$ metals at the step-edge  of a $5d$ substrate. But it is also a system in its own right. Recently, the magnetic properties of various bi-metallic 3$d$-5$d$ chains of linear and zigzag shape have been investigated.\cite{Tung2011,che-2,Negulyaev_13}

The calculations are carried out within the full-potential linearized augmented plane wave method (FLAPW)\cite{FLAPW-1,FLAPW-2} as implemented in the {\tt FLEUR} code\cite{fleur}. In order to deal with the large unit cell anticipated for chiral magnetic spirals, we treat the magnetic structure in reciprocal space by making use of the generalized Bloch theorem\cite{Herring:66.1,Sandratskii:86.2,Sandratskii:91.1} in the absence of the spin-orbit interaction, which allows the calculation of incommensurate magnetic spirals in the chemical unit cell. The  spin-orbit interaction is then treated in first-order perturbation theory in the basis of the spin-spiral solutions. The MAE is determined by separate calculations and all results are discussed in terms of the model Hamiltonians for the different spin-interactions (viz., Heisenberg, DMI and MAE).
 
Our findings show that without SOI all systems are ferromagnets with the exception of the Fe-Pt and Co-Pt bi-atomic chains. For these two chains, we expect a magnetic exchange spiral that is degenerate with respect to the right- or left rotation sense. Including the spin-orbit interaction we find that the hard magnetization axis is normal to the plane of the zig-zag chain and thus any spiral should be of cycloidal nature where the magnetization rotates in the plane of the zig-zag chain. The DMI depends critically on the substrate, {\it i.e.}\ the $5d$ atom of  the bi-atomic zig-zag chain, the sign of the DMI flips each time, when moving from Ir, to Pt and then to Au. Among all chains, for the Fe-Pt and Co-Pt chains, the DMI is sufficiently strong to stabilize a chiral magnetic ground state of left-rotational sense.

Surprisingly little is known about the relation between the DMI interaction and the electronic orbitals that contribute to it. The rather clear nature of the electronic structure of the bi-atomic zig-zag chain invites the development of a minimal tight-binding model consisting of four relevant $d$-orbitals located at two $3d$ atoms and one $5d$ atom arranged in a triangular geometry. In this paper, we present the results of  this simple tight-binding model, which represents the essential features of the problem elucidating the factors controlling the sign and strength of DMI in these 3$d$-5$d$ transition-metal zigzag chains.  

The paper is organized as follows: Sec.~\ref{sec-com} describes the computational methodology required for determination of the DMI and MAE from first-principles calculations. In Sec.~\ref{Results}, the results for the 3$d$-5$d$ bi-atomic chains are presented. In Sec.~\ref{sec-tb} we describe the tight-binding model for the trimer in detail and from the results we draw analogies to the infinite 3$d$-5$d$ chains. Finally, we conclude our findings in Sec.~\ref{sec-sum}.

\section{Computational methodology}\label{sec-com}

\subsection{Structural optimization}

We have modeled free-standing planar zigzag bi-atomic chains of 3$d$-5$d$ elements, as shown in Fig.~1. For the calculations, we have used the film version of the full-potential linearized augmented planewave (FLAPW) method as implemented in the J\"{u}lich DFT code {\tt FLEUR}.\cite{fleur} For our one-dimensional structures, we choose a large rectangular two-dimensional unit cell of 20 a.u.\ along the $y$-direction to minimize the interaction between periodically repeated images of  one-dimensional infinite chains containing one $3d$ and one $5d$ atom. Then, we optimize the lattice parameter $a$ for the magnetic chains, corresponding to the unit-cell length in $x$-direction, and the bond length $d$, by carrying out spin-polarized calculations applying the revised Perdew-Burke-Ernzerhof (rPBE)\cite{Zhang} exchange correlation  functional within the Generalized Gradient Approximation (GGA). Our unit cell is perfectly embedded into two semi-infinite vacua in the $\pm z$-directions. The muffin-tin sphere around each atom was chosen to be $2.2$~a.u.\ for all chains. A careful convergence analysis shows that a plane-wave cutoff of $3.8~\mathrm{a.u.}^{-1}$ and 48 ${k}$-points along the positive half-space of the one-dimensional Brillouin-zone are sufficient to obtain converged structural parameters $a$ and $d$ in non-magnetic calculations. For completeness, we mention that in our set-up, the inversion symmetry is broken due to a lack of reflection symmetry along the $xz$-plane.

\begin{figure}
    \centering
      \includegraphics[width=2.5in]{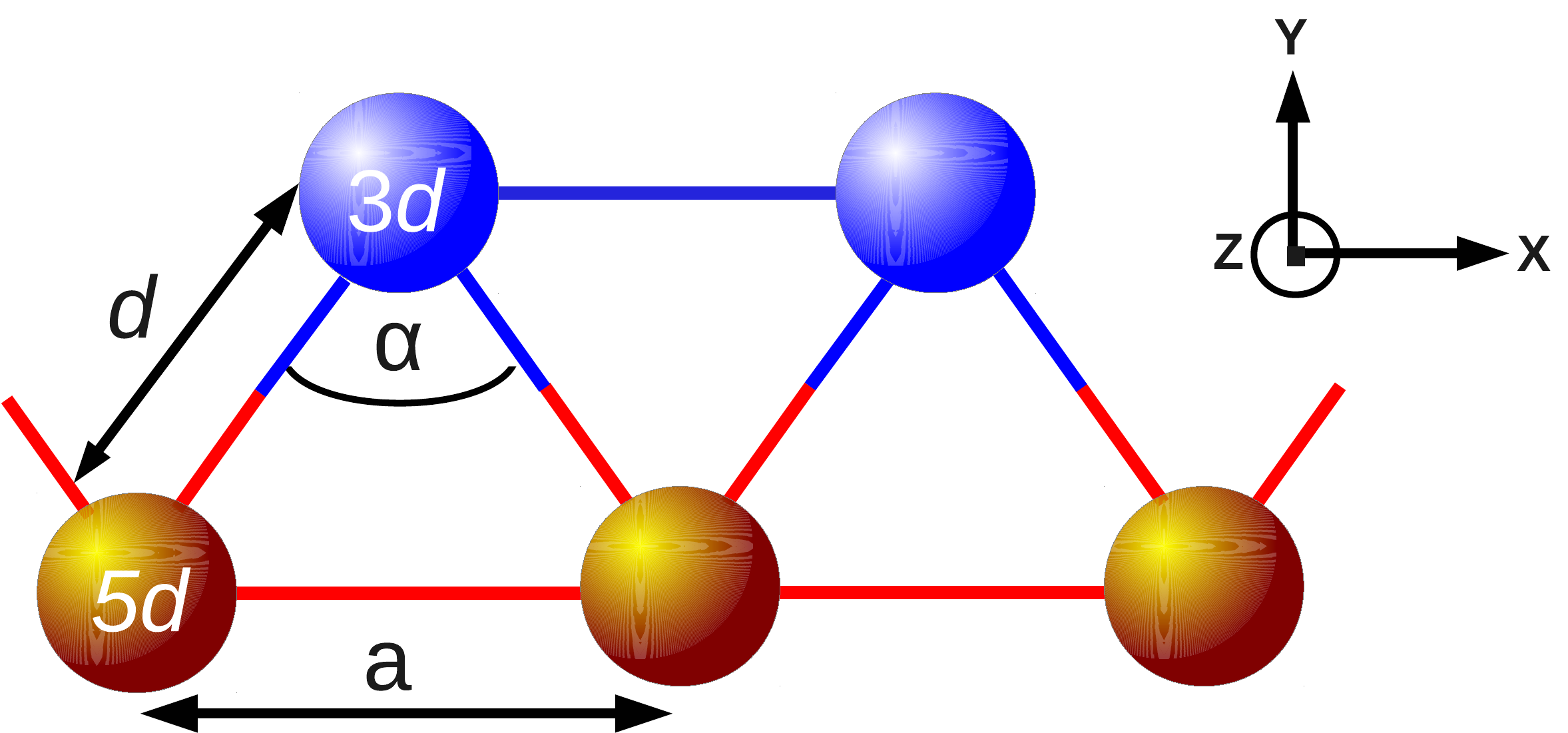}
     \caption{(Color online) Structure of the 3$d$-5$d$ transition metal chains. The lattice parameter $a$ denotes the equilibrium bond length between two consecutive 3$d$ (5$d$) atoms. $d$ represents the distance between the 3$d$-5$d$ atoms and $\alpha$ is the angle spanned by the 5$d$-3$d$-5$d$ atoms.}
    \label{model3d}
\end{figure} 

\subsection{Collinear magnetic calculations}
Using the optimized geometry, we calculated the energy difference between collinear states (ferromagnetic and antiferromagnetic, respectively) with 48 ${k}$-points, using the exchange correlation functional GGA-rPBE  on the one hand, and the Vosko, Wilk, and Nusair (VWN) functional\cite{Vosko} within the local density approximation (LDA) on the other. All magnetic interactions are provided as calculated by the  LDA functional as experience has shown that it gives a more realistic description of the magnetic interaction energies.

\subsection{Spin-spiral calculations}
We consider flat, homogeneous spin spirals, which are defined by two quantities: the spin-spiral wave vector $\mathbf{q}$ and a rotation axis. The former has three properties: (i) the direction of $\mathbf{q}$ corresponds to the propagation direction of the spin spiral (in our case it is limited to the $x$-direction due to the one-dimensional nature of the chains and we omit the vector character of $\mathbf{q}$ in the following), (ii) its magnitude determines the wave length $\lambda = 2 \pi \, \lvert q \lvert ^{-1}$ of the spin spiral, and finally (iii) the sign of $q$ determines the rotational sense of the spin spiral. If $q > 0$ ($q < 0$) we refer to a counter-clockwise (clockwise) or left (right) rotating spiral. 
To finalize the definition of the spin spirals, we comment that for `flat' spin spirals all magnetic moments are rotating in one plane perpendicular to the rotation axis. There are two special q-points in the one-dimensional Brillouin-zone that deserve mentioning, $q=0$ that represents the ferromagnetic alignment, and $q = \pm 0.5 ~ \frac{2\pi}{a}$ that represents antiferromagnetic alignment.

We have performed self-consistent total-energy calculations of spin spirals within the scalar-relativistic approximation ({\it i.e.}\ without SOI) using both, the GGA-rPBE and LDA-VWN exchange-correlation functionals. In this case, we can without loss of generality choose a rotation-axis along the $z$-direction and exploit the generalized Bloch theorem,\citep{Herring,sandratskii,sandratskii2} which allows for a calculation of spin spirals in the chemical unit cell rather than a possibly large supercell and thus reduces the computational effort considerably. In a second step, we have estimated the effect of SOI on the spin-spiral energies in first-order perturbation-theory (cf.\ Sec.~\ref{SecMethDMI}). For all spin-spiral calculations a dense $k$-mesh of 384 $k$-points has been used.

Let us first look at the case without SOI: the corresponding interaction energy between magnetic moments can be described in terms of a Heisenberg model as
\begin{equation}
  E_0(q) = \sum_{i,j} J_{ij}~ \mathbf{S}_{i} \cdot \mathbf{S}_{j}, \label{heis}
\end{equation}
where the direction vector of the magnetic moment, $\mathbf{S}_j$, at lattice site $j$ is parameterized by the magnetic spiral $\mathbf{S}_{j} = \left( \cos(j \, a \,q), \sin(j\, a \,q), 0 \right)^\mathrm{T}$,
and the sign of the isotropic exchange integrals $J_{ij}$ determines whether the magnetic interaction between the sites $i$ and $j$ is ferromagnetic ($J<0$) or antiferromagnetic ($J>0$). Non-trivial spin-spiral ground states can be formed, if the interaction between different neighbors is competing in sign and strength in a form that the mutual exchange interaction is frustrated. Such spirals are exchange spirals in opposite to chiral spirals induced by the DMI. Exchange spirals are achiral in the sense that the energies are degenerate with respect  $q$ and $-q$, which is reflected by the dot-product of the Heisenberg model.

\subsection{Calculation of Dzyaloshinskii-Moriya interaction} \label{SecMethDMI}

When considering the SOI for a spin-spiral state, two more energy contributions appear: A magneto-crystalline anisotropy energy (MAE, cf.\ Sec.~\ref{SecMethMCA}) and the Dzyaloshinskii-Moriya interaction, which in terms of a spin model is of the form
\begin{equation}
  E_{\rm DM}(q) = \sum_{i,j} \textbf{D}_{ij} \cdot \left( \textbf{S}_{i} \times \textbf{S}_{j} \right).
\end{equation}
Here, the antisymmetric exchange constants $\textbf{D}_{ij}$ are called Dzyaloshinskii-Moriya (DM) vectors, which determine the strength and sign of DMI. Due to the cross-product between the two magnetic moments, canted spin-structures of a particular handedness are favored by this energy term. The type of handedness depends on the sign of the DM vectors with respect to the spin-rotation axis.  As a result, the degeneracy of spin spirals with respect to the direction of the rotation axis is lifted: $E_{\rm DM}$ will become extremal for a rotation axis parallel to the DM vector. For the zigzag chains investigated in this work,  the $xy$-plane is a global mirror plane (${\cal M}: (S_x, S_y, S_z) \rightarrow (-S_x, -S_y, S_z)$),  and through plain symmetry arguments the DM vector $\textbf{D}=(0,0,D)$ is pointing along the $\pm z$-direction, thus preferring flat spin spirals with a rotation in the $xy$-plane. Within our geometry, we define the chirality index $C = \mathrm{\mathbf{e}_z} \cdot \left( \textbf{S}_i \times \textbf{S}_{i+1} \right)$ and call the magnetic structure left-handed (right-handed) for $C=+1$ ($C=-1$).

For the calculation of the energy of spin spirals including the SOI, the generalized Bloch theorem cannot be applied any more, because atoms with different directions of the magnetization can be distinguished by their spin-orbit interaction energy. One possible way could be to use large supercells in which the magnetic structure is commensurate, to large computational costs. However, due to the much smaller SOI energy compared to the total energy of the spin spiral,\citep{Heide,marcusthesis} we treat SOI as a perturbation to the system. This allows us to find the energy levels and the wave functions of the unperturbed system, $\epsilon^0_{k\nu}(q)$ and $\psi^0_{k\nu}(q,{\bf r})$, for the one-dimensional Bloch vector $k$ and band index $\nu$, using the chemical unit cell only. Then we estimate the shift $\delta \epsilon_{k\nu}$ of these levels due to the SOI Hamiltonian $\mathcal{H}_{\rm SO}$, in the basis of spin-up and spin-down states as
\begin{equation} 
 \delta \epsilon_{k\nu} = \begin{pmatrix} \langle \psi^{(\uparrow)}_{k\nu} \rvert & \langle \psi^{(\downarrow)}_{k\nu} \rvert \end{pmatrix}
                         \begin{pmatrix} \mathcal{H}_{\rm SO}^{(\uparrow,\uparrow)} & \mathcal{H}_{\rm SO}^{(\uparrow,\downarrow)} \\ 
                                         \mathcal{H}_{\rm SO}^{(\downarrow,\uparrow)} & \mathcal{H}_{\rm SO}^{(\downarrow,\downarrow)} \end{pmatrix}
                         \begin{pmatrix} \lvert \psi^{(\uparrow)}_{k\nu} \rangle \\ \lvert \psi^{(\downarrow)}_{k\nu} \rangle \end{pmatrix}.
\label{delta_e_abintio}
\end{equation}
Summing up these energy shifts over all occupied states of the unperturbed system, we find an energy correction corresponding to the Dzyaloshinskii-Moriya interaction,
\begin{equation}
  E_{\rm DM}(q) = \sum_{k\nu}^{\rm occ.} \delta \epsilon_{k\nu}(q). \label{EsumSOC}
\end{equation}
Because each level exhibits the symmetry $\delta \epsilon_{k\nu}(-q) = - \delta \epsilon_{k\nu}(q)$, this antisymmetric behavior will be inherited by the sum, $E_{\rm DM}(-q) = -E_{\rm DM}(q)$, and only the spin spirals with positive $q$ must be computed. Obviously,  $E_{\rm DM}(q)$ is an odd function of $q$ and for small $|q|$, $E_{\rm DM}(q)\simeq Dq$, where $D$ takes the role of an effective DM vector in $z$-direction and is a measure for the strength of the DMI. The Dzyaloshinskii-Moriya interaction in the 3$d$-5$d$ chains was calculated using a LDA-VWN functional and with a dense mesh of $384$ $k$-points.

\subsection{Magneto-crystalline anisotropy} \label{SecMethMCA}

The second energy contribution due to SOI is the magneto-crystalline anisotropy energy (MAE). It will generally tend to a collinear alignment of magnetic moments along the easy axis of the system, and thus competes against the DMI. To be more precise, it will compete against any non-collinear magnetic structure, since then there are always magnetic moments pointing away from the easy axis that will increase the energy of the system. Based on our results (as discussed in Sec.~\ref{Results}), we can focus the  following discussion on an easy axis ({\it i.e.}\ the direction with lowest energy) that is either in the $x$ or $y$ axis. Then, let $K_{1}$ be the difference between the energies of these two directions, and let $K_{2}$ be the difference between the energies along the $z$-axis and the easy axis. Any homogeneous, flat spin spiral rotating in the $xy$-plane will then have an average MAE per atom of $\frac{1}{2}K_{1}$.

In order to compute the MAE, we have performed collinear ({\it i.e.}\ ferromagnetic) calculations, where we chose the magnetic moments to be fixed along the three crystal axes, $x$, $y$ and $z$, respectively. The spin-orbit interaction was included self-consistently in our calculations, using 192 $k$-points in the whole Brillouin zone. By comparing the total energies of the three calculations, we obtain values for $K_{1}$ and $K_{2}$.

\subsection{Ground state formation and inhomogeneity} \label{SecMethGroundState}

Considering these three energy contributions, a chiral homogeneous spin spiral with wave vector $q$ will be established by the DMI out of the ferromagnetic state, {\it i.e.\ } $q=0$, only if their sum yields an energy lower than the  ferromagnetic state, {\it i.e.\ } $E_0(q) + E_{\rm DM}(q) + K_1/2 < E_0 (q=0)$.

Although the {\it ab initio} calculations impose  homogeneous spin spirals, the possible formation of inhomogeneous spirals can be analysed on the basis of a micromagnetic model of one-dimensionally spiralling  magnetic structures developed by Dzyaloshinskii\cite{Dzyaloshinsky:65:2} with micromagnetic parameters deduced from the homogeneous calculations. In homogeneous spin spirals the angle of the magnetization direction changes by the same amount from atom to atom. If the magnetic anisotropy is strong, it seems natural that the magnetization direction along the easy axis is preferred and we expect small angles of rotation in the vicinity of the easy axis accompanied by fast rotations into the hard axes and back. Within this micromagnetic theory the degree of inhomogeneity is quantified by a parameter\cite{Heide:2011:micmod} $\kappa= 16/\pi^2 \cdot A K_1/D^2$, whereas the micromagnetic parameters are taken from a fit of a quadratic energy form $E = A q^2 + D q + K_1/2$ to the {\it ab initio} energy dispersion in the vicinity of the energy minimum. If $\kappa=0$ the spiral is perfectly homogeneous, for $\kappa=1$ the spiral separates into two collinear domains, separated by a chiral domain wall.

\section{Results and discussion}\label{Results}

\subsection{Structural properties and magnetic moments}

\begin{figure*}
  \centering
  \includegraphics[angle=270,width=4.8in]{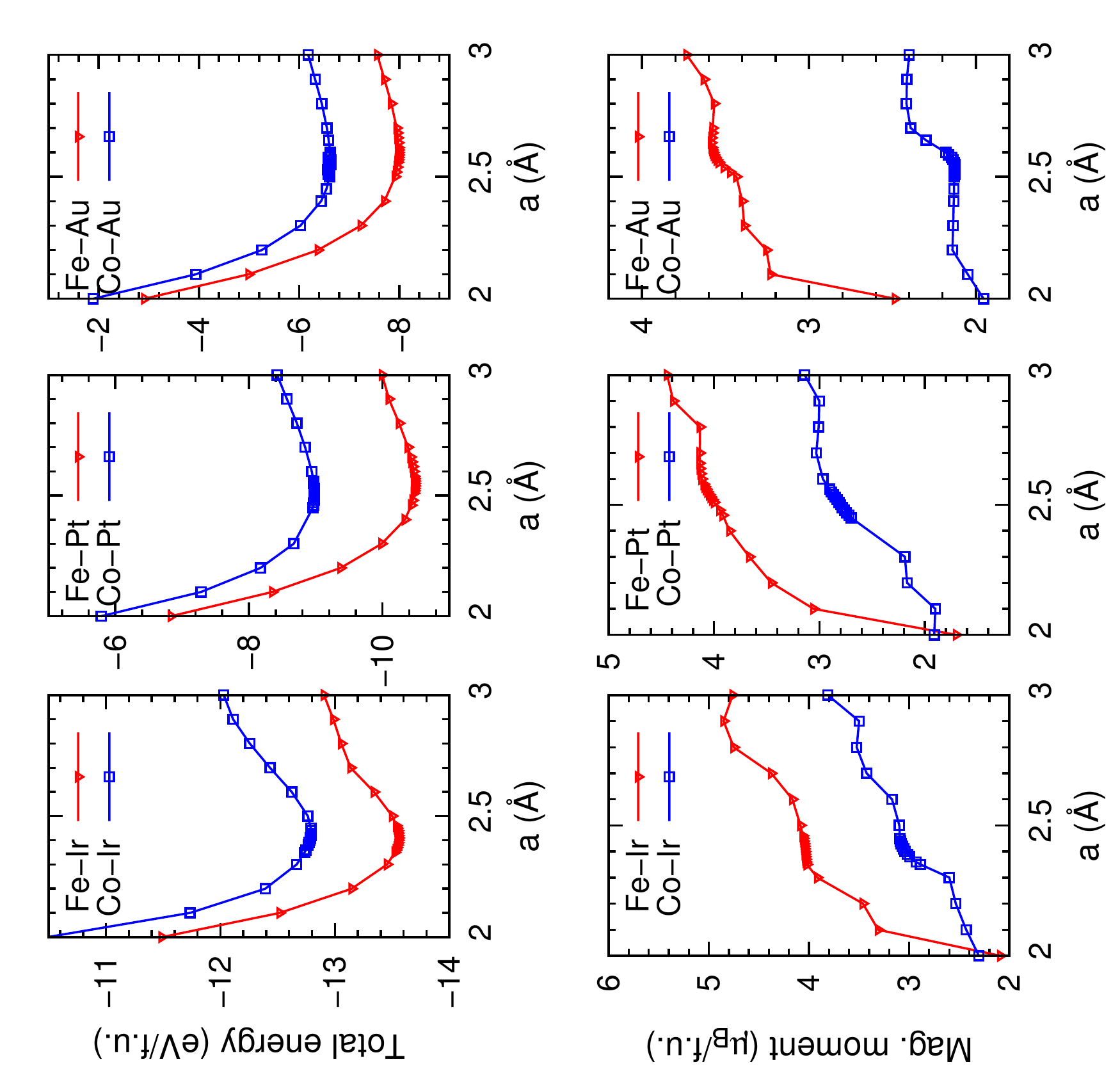}
 \caption{(Color online) Total energies (upper panel) and magnetic moments (lower panel) of Fe- and Co-$5d$ chains plotted as a function of lattice constants $a$ ({\it i.e.}\ $E(a, d_o)$ and $M(a, d_o)$) for optimal $d_o$ at given $a$.  The results for Fe-5$d$ chains are shown by red lines with triangles, whereas the blue lines with squares denote results for Co-5$d$ chains.}
  \label{binding}
\end{figure*}

\begin{table}
 \begin{tabular} {c  c  c  c c  c  c c c c} \hline \hline
  System & & $a_o$ (\AA) &  & $d_o$ (\AA) & &  $\alpha_o$ (in $^{\circ}$)  &  & \multicolumn{2}{c}{$M$($\mu_{B}$/atom)}\\ \cline{9-10}
         & &         &  &           & &                            &  & 3$d$ & 5$d$  \\
  \hline
  \hline
  Fe-Ir & & 2.41 & & 2.35 & & 61.83 & & 3.11 & 0.67 \\
  Fe-Pt & & 2.55 & & 2.44 & & 63.08 & & 3.27 & 0.53 \\
  Fe-Au & & 2.62 & & 2.59 & & 60.76 & & 3.26 & 0.08 \\
  \hline 
  Co-Ir & & 2.45 & & 2.37 & & 62.07 & & 2.06 & 0.68 \\
  Co-Pt & & 2.50 & & 2.43 & & 62.02 & & 2.18 & 0.34 \\
  Co-Au & & 2.59 & & 2.57 & & 60.42 & & 2.15 & 0.07 \\
  \hline
 \end{tabular}
 \vskip 0.1in
\caption{Structural parameters of the optimized zigzag 3$d$-5$d$ metallic chains. $a_o$ is the lattice constant, $d_o$ represents the 3$d$-5$d$ distance and $\alpha_o$ is the angle between the 5$d$-3$d$-5$d$ atoms (cf.\ Fig.~\ref{model3d}). The magnetic moment $M$ on the 3$d$ and 5$d$ atoms are also listed.}
 \label{table-binding}
\end{table}

In the present paper, we have investigated zigzag 3$d$-5$d$ bi-metallic chains as shown in Fig.~\ref{model3d}. We observed, that all chains give well defined, unique minima in the total energy with respect to the lattice constant $a$ (upper panel of Fig.~\ref{binding}).
Table~\ref{table-binding} shows the optimized geometrical properties as well as the magnetic moments of the two kinds of atoms. The values of $a$ and $d$ indicate that isosceles triangles are formed, similar to gold\cite{vikas1} and nickel\cite{vikas2} zigzag chains.
The magnetic moment of 3$d$ atoms is larger in  Fe-5$d$ as compared to Co-5$d$ chains. The $5d$ atoms show relatively small induced magnetic moments and they depend weakly on the choice of the 3$d$ atoms. The induced magnetization is larger for 5$d$ atoms with smaller atomic number ($Z$). 

We have also investigated the variation of the total magnetic moment in the unit cell \footnote{The total magnetic moment is a sum of the magnetic moments per atom, reported in Table~\ref{table-binding}, and the magnetization in the interstitial region.} as a function of the lattice constant $a$ (lower panel of Fig.~\ref{binding}). As before, the magnetic moments for Fe-5$d$ chains are larger than those of Co-5$d$ chains for a large range of $a$. With increasing $a$, the magnetic moment increases as the electron wave functions tend to become more atomic. This variation in the magnetic moment in Fe-5$d$ chains is larger than that of Co-5$d$ chains. In addition, the variation in the magnetic moment decreases as the lattice constant is increased in all 3$d$-5$d$ chains. As the lattice constant decreases close to 2\AA, the  magnetic moment of Fe-5$d$ chains decreases sharply, indicating the possibility of a magnetic transition during compression.

\subsection{Isotropic exchange interaction}
\label{chap_isotropic_exchange}

\begin{figure*}
    \centering
    \includegraphics[angle=270,width=5.7in]{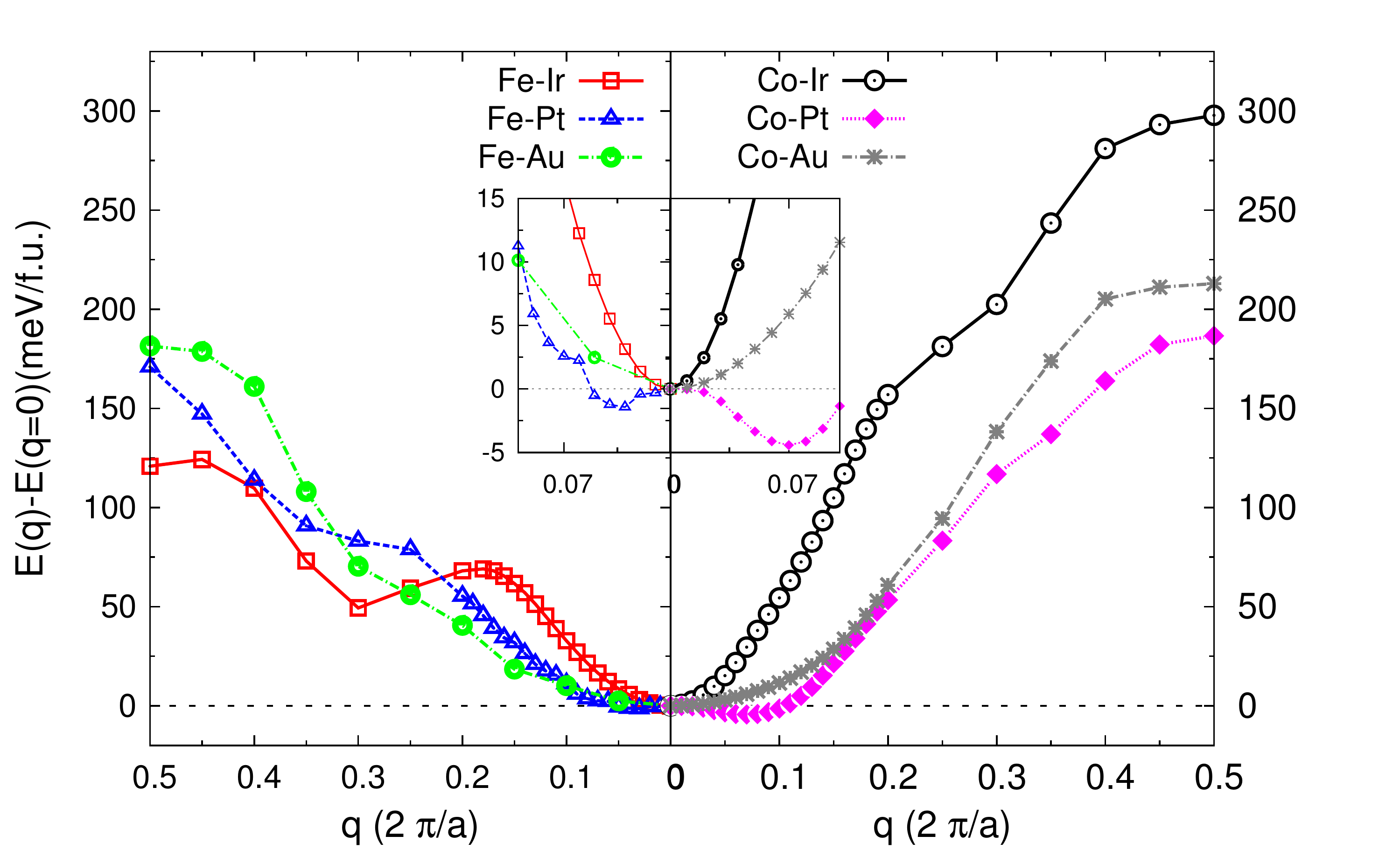}
   \caption{(Color online) The dispersion energy of flat spirals for Fe-5$d$ (left) and Co-5$d$ chains (right) without SOI as a function of the spin-spiral vector $q$ calculated using the LDA-VWN functional. Because of the symmetry $E_0{(q)}=E_0(-q)$, we show only the dispersion energies for $q>0$. The inset shows the magnified region near the ferromagnetic state ($q=0$).}
  \label{spinspiral}
\end{figure*} 

In Fig.~\ref{spinspiral}, we present the calculated energies of flat, homogeneous spin spirals of 3$d$-5$d$ chains in the scalar-relativistic approximation (not considering SOI). In this case, the dispersion energy is an even function of the spin-spiral vector ({\it i.e.}, $E_{0}(q)=E_{0}(-q)$). Our results demonstrate, that the ferromagnetic state is energetically most stable in most 3$d$-5$d$ chains, except for Fe-Pt and Co-Pt, which show a (non-collinear) spin-spiral ground state. In Co-Pt, the energy at $q=\pm 0.07$ (henceforth the values of $q$ will be given in units of $2\pi/a$, which corresponds in this case to an angle of $25.20^{\circ}$ between adjacent unit cells) is lower than that of the ferromagnetic state by 4.4 meV/f.u. In Fe-Pt, the energy minimum at $q= \pm 0.03$ (corresponding to an angle of $10.79^{\circ}$) is only $1.4\,\mathrm{meV/f.u.}$ lower than the ferromagnetic state.

In Co-Ir and Co-Au chains, the  spin-spiral dispersion energies presented in Fig.~\ref{spinspiral} show a typical parabolic behavior around the ferromagnetic ($q=0$) and antiferromagnetic ($q=0.5$) states. For Fe-based chains and the Co-Pt chain, the shape of the dispersion energies  show deviations from the pure cosine behavior with  a minimum value at $q=0$ and a maximum value at the zone-boundary. Obviously, exchange interactions $J_{ij}$ between more distant neighbors become important.  For example, a dip around $q \approx \pm 0.3$ is observed in the Fe-$5d$ chains, with Fe-Ir having a more pronounced manifestation. However, these longer-distant interactions do not influence the magnetic ground state of these chains. Instead, the further-distant interactions in the Pt-based chains have an influence on the magnetic ground state. They compete with the ferromagnetic nearest neighbor interaction between the 3$d$ atoms, in total leading to a local minimum at small q-values. As result we obtain for these systems an incommensurable spin-spiral ground state at $q=\pm 0.035 $ for Fe-Pt and   $q=\pm 0.07 $ for Co-Pt.  

\begin{table}
\resizebox{3.4in}{!}{
\begin{tabular} {c  c  c c  c c  c } \hline \hline
  System  &  Collinear  & Spin-Spiral &  Collinear & Spin spiral \\
          & (GGA-rPBE)  & (GGA-rPBE) &  (LDA-VWN)   & (LDA-VWN)   \\
  \hline
  \hline
  Fe-Ir & \phantom{0}92  & 105 & 117  & 121 \\
  Fe-Pt &  112 & 126 & 162 &  170 \\
  Fe-Au &  119 & -- & 185  & 181 \\
  \hline
  Co-Ir &  154  & -- & 294  & 296 \\
  Co-Pt &  146 & 155  & 182 & 186 \\
  Co-Au & 171 & 168  & 212 & 213 \\
  \hline
 \end{tabular}}
 \vskip 0.1in
\caption{(Color online) Energy difference in meV/f.u.\ between ferromagnetic and antiferromagnetic states for 3$d$-5$d$ chains. The collinear results were obtained using a supercell approach, whereas the spin-spiral calculations were performed exploiting the generalized Bloch theorem. Some of the GGA spin-spiral calculations did not converge.}
 \label{Table-SS}
\end{table}

For completeness, we compared for all chains the energy difference between the ferromagnetic (FM) and antiferromagnetic (AFM) configuration evaluated using the GGA-rPBE exchange-correlation functional with the difference obtained by the LDA-VWN functional for two types of calculations, one carried out by the spin-spiral formalism and one by collinear calculations. All energy differences have been evaluated for the ground-state geometry obtained by the GGA functional.  From Table~\ref{Table-SS} it can be seen that the LDA-VWN  functional gives significantly larger (25\% to 50\%)  energy differences.  The spin-spiral calculation and the collinear calculation for the antiferromagnetic configuration are different in one respect in that the quantization axes of the $5d$ atoms are rotated by $90^\circ$ with respect to the one of the $3d$ atoms for the spin-spiral but are parallel for the collinear calculations. However, the magnetic moment of the 5$d$ atom is much reduced in the antiferromagnetic state due to frustration, and in turn the direction of the quantization axis has little influence on the total energy for the AFM. The frustration occurs because the moment of the 5$d$ atom couples ferromagnetically to the moment of the 3$d$ atoms (\textit{cf.}~Table~\ref{table-binding}), and for an antiferromagnetic configuration any finite moment of the 5$d$ atom would be parallel to the moment of the one 3$d$ atom and antiparallel to the moment of the other 3$d$ atom. The energy difference is in general larger for GGA results (14~meV/f.u.\ for Fe-Pt and 9~meV/f.u.\ for Co-Pt) than for LDA results (8~meV/f.u.\ for Fe-Pt and 4~meV/f.u.\ for Co-Pt). Thus, the total energy depends only very little on the choice of the quantization axis of the $5d$ atoms and a further optimization of the direction of this axis is not necessary.

\subsection{Effect of spin-orbit interaction on magnetism}

\subsubsection{Magneto-crystalline anisotropy energy}
The magneto-crystalline anisotropy energy (MAE) is extracted from the total-energy calculations of ferromagnetic states, with the magnetic moments pointing along the three high-symmetry directions. We find that for all investigated chains the $z$-axis is the hard-axis, $K_{1}<K_{2}$ (cf.\ Sec.~\ref{SecMethMCA}) and subsequently, for all  chains the easy axis lies  in the $xy$-plane  (cf.\ Fig.~\ref{model3d}) of the bi-atomic chains (Table~\ref{Table-MAE}). The Fe-5$d$ chains, except for Fe-Au,  prefer an uniaxial magnetization along the $x$-axis. In contrast, the Co-5$d$ chains prefer the $y$-axis as the easy axis, except for Co-Ir. 
The 3$d$-Pt chains exhibit the smallest in-plane anisotropy $K_{1}$, whereas, 3$d$-Ir chains have a very large $K_{1}$. The 3$d$-Au chains have the smallest out-of-plane anisotropy. This is a consequence of the hybridization between the spin-split transition-metal  3$d$ states with the spin-orbit-interaction carrying 5$d$-states. This hybridization is smaller for Au than for Pt or Ir, because the 5$d$-states of Au are 3 eV below the Fermi energy, while for Pt and Ir the $5d$ states are crossing the Fermi energy and then the interaction of magnetism with SOI is much stronger.  In general, the magneto-crystalline anisotropy energies found here are large compared to the values found for typical bulk structures,\cite{gay,dorantes,Szunyo}
as expected for systems with reduced dimensions.

\begin{table}
 \begin{tabular} {c  c  c  c  } \hline \hline
  System & \hskip 0.2in Easy Axis & \hskip 0.2in $K_{1}$  (meV/f.u.) & \hskip 0.2in $K_{2}$  (meV/f.u.) \\
  \hline \hline
  Fe-Ir & $x$ & 2.4 & 5.3 \\
  Fe-Pt & $x$ & 0.8 & 7.3 \\
  Fe-Au & $y$ & 1.8 & 2.1 \\
  \hline
  Co-Ir & $x$ & 15.8 & 20.7 \\
  Co-Pt & $y$ & 0.2  & 12.3 \\
  Co-Au & $y$ & 1.4  & 2.1  \\
  \hline
 \end{tabular}
 \vskip 0.1in
 \caption{Magneto-crystalline anisotropy energy in meV/f.u. $K_{1}$ represents the energy difference between the easy axis and its perpendicular conjugate in the plane of the chain. $K_{2}$ is the energy difference between the easy axis and $z$ axis.}
 \label{Table-MAE}
 \end{table}

We also calculated the shape anisotropy due to the classical magnetic dipole-dipole interactions, using the magnetic moments listed in Table~\ref{table-binding}. For all chains, we observed an energetically most favorable direction of the magnetic moments along the wire axis. However, the magneto-crystalline anisotropy due to spin orbit coupling is 2-3 orders of magnitudes larger than the shape anisotropy and hence dominates the magnetic anisotropy contribution to the energy of the system.

\subsubsection{Dzyaloshinskii-Moriya interaction energy}  \label{DMI_abinitio}

\begin{figure*}
  \centering
  \hspace{0.5cm} {\large (a)} \hspace{7.2cm} {\large (d)}\\
    \includegraphics[angle=270,width=2.8in]{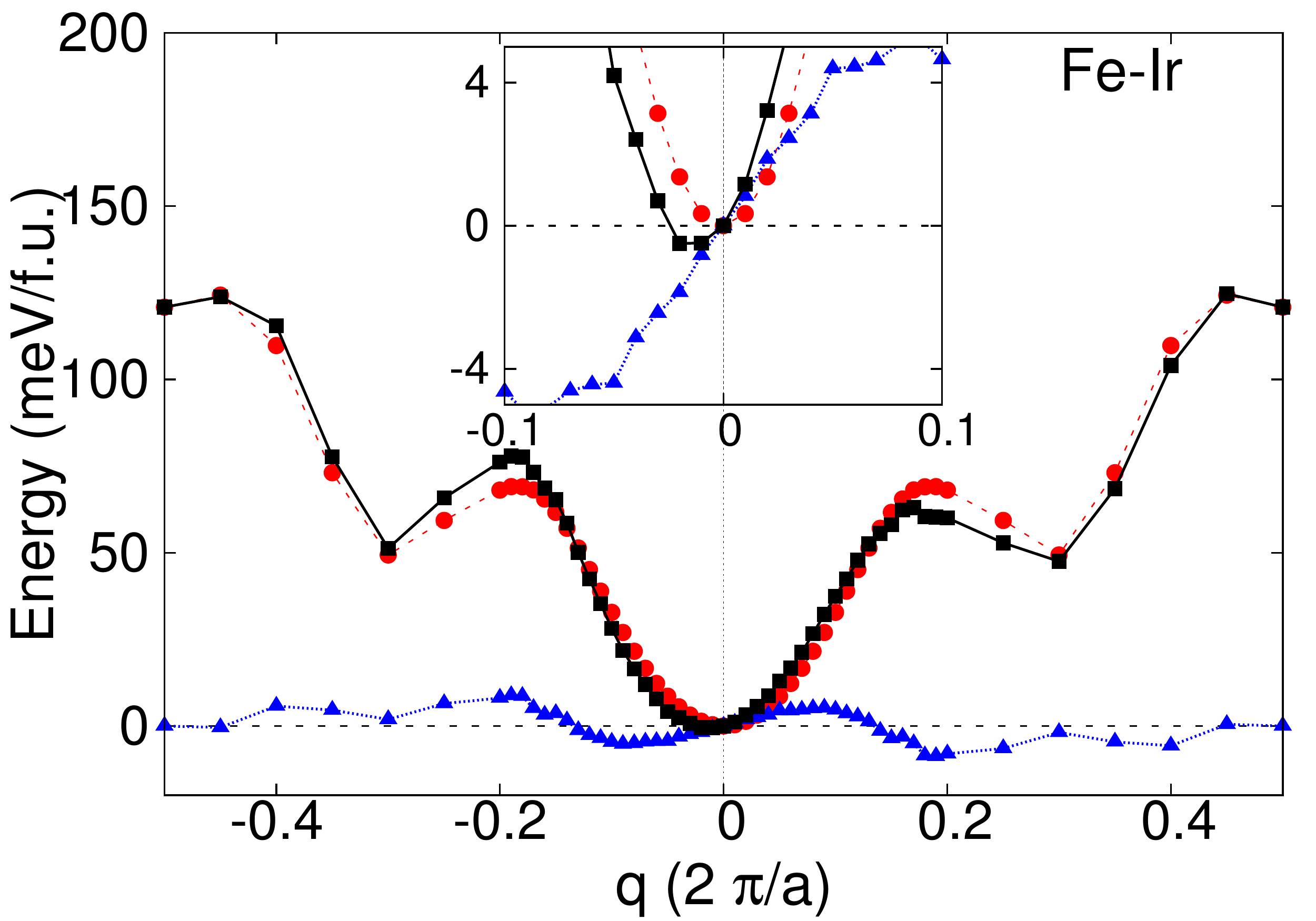}\hskip 0.3in \includegraphics[angle=270,width=2.8in]{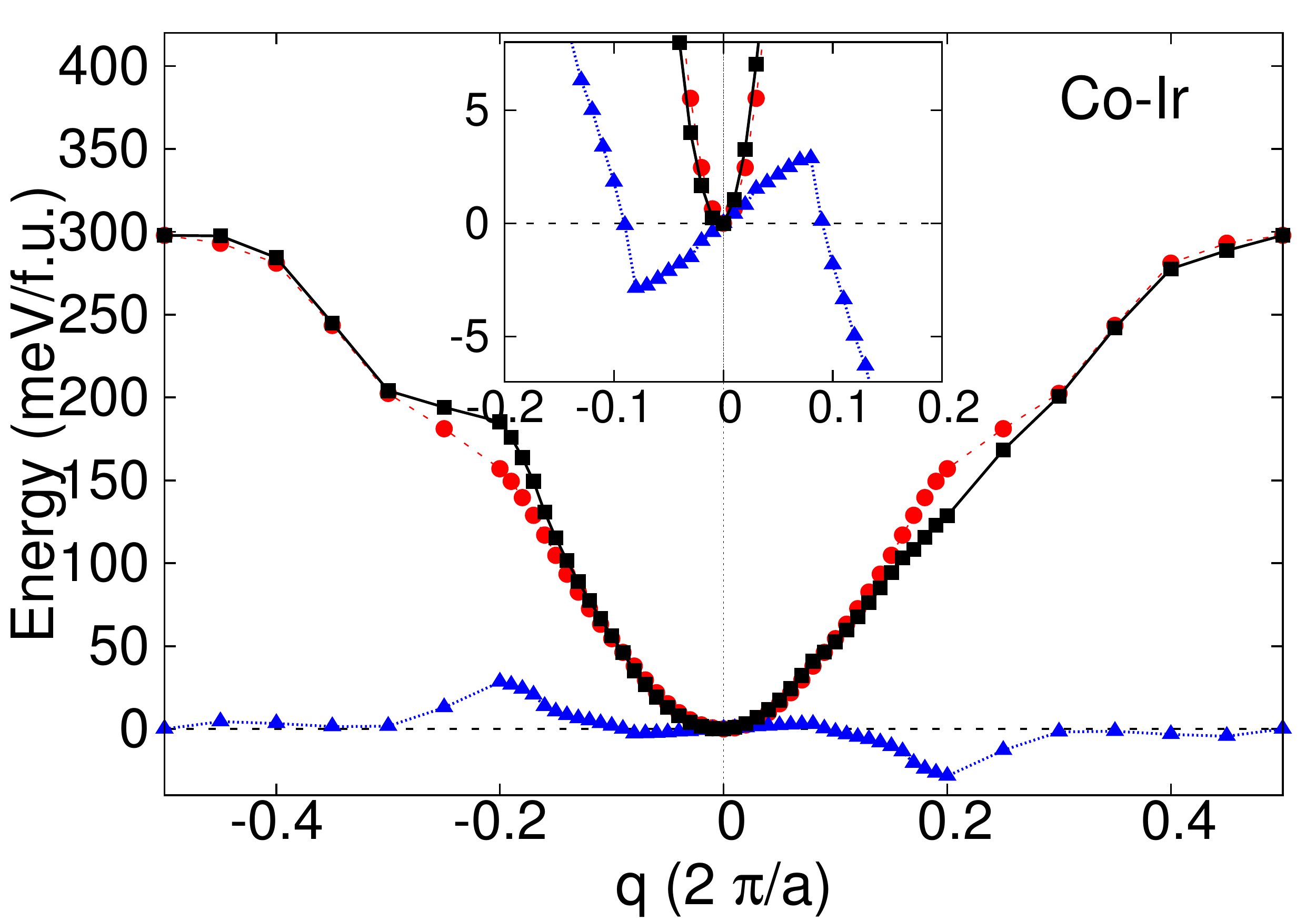}\\
       \vskip 0.1in
   \hspace{0.5cm} {\large (b)} \hspace{7.2cm} {\large (e)}\\
    \includegraphics[angle=270,width=2.8in]{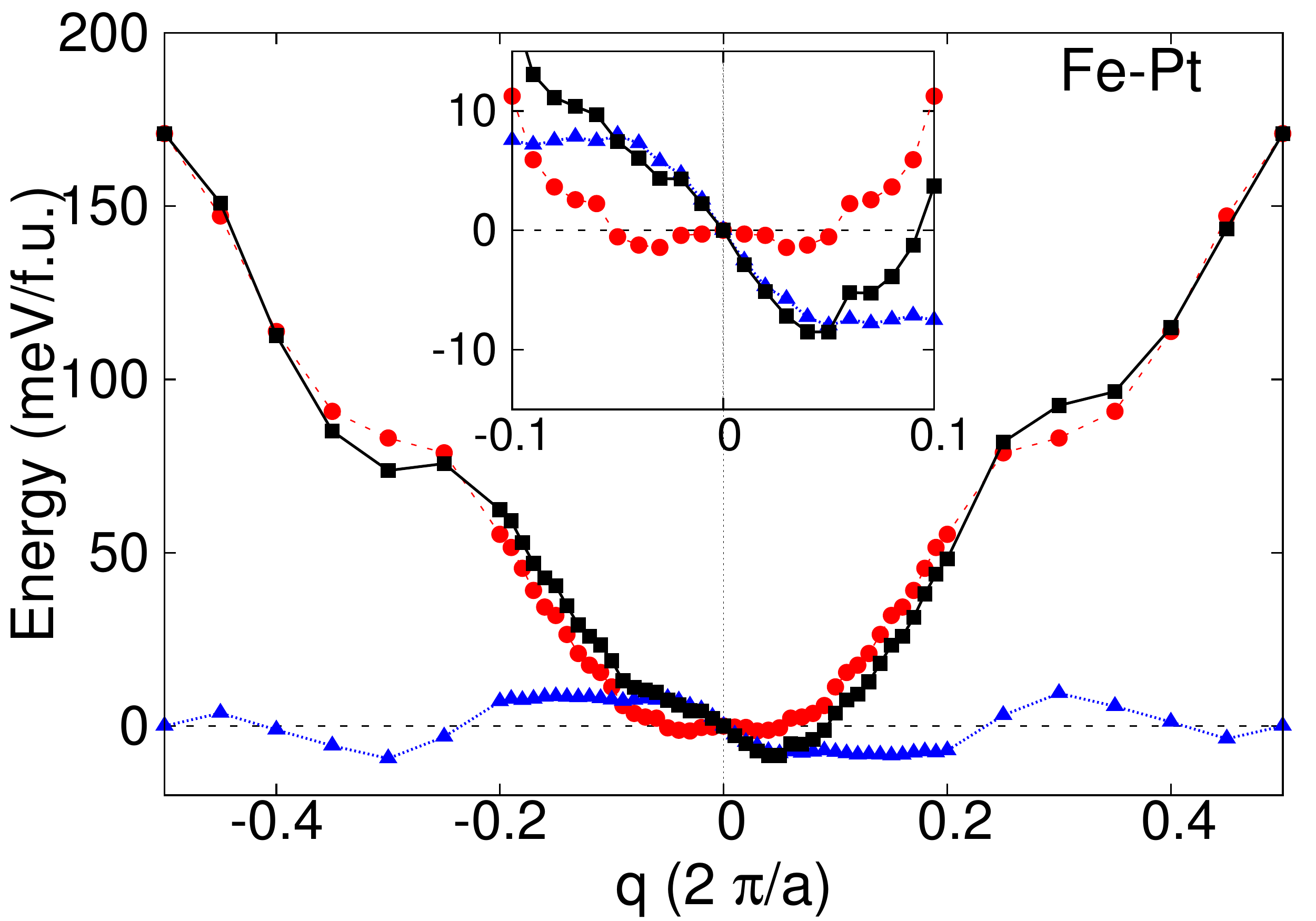}\hskip 0.3in \includegraphics[angle=270,width=2.8in]{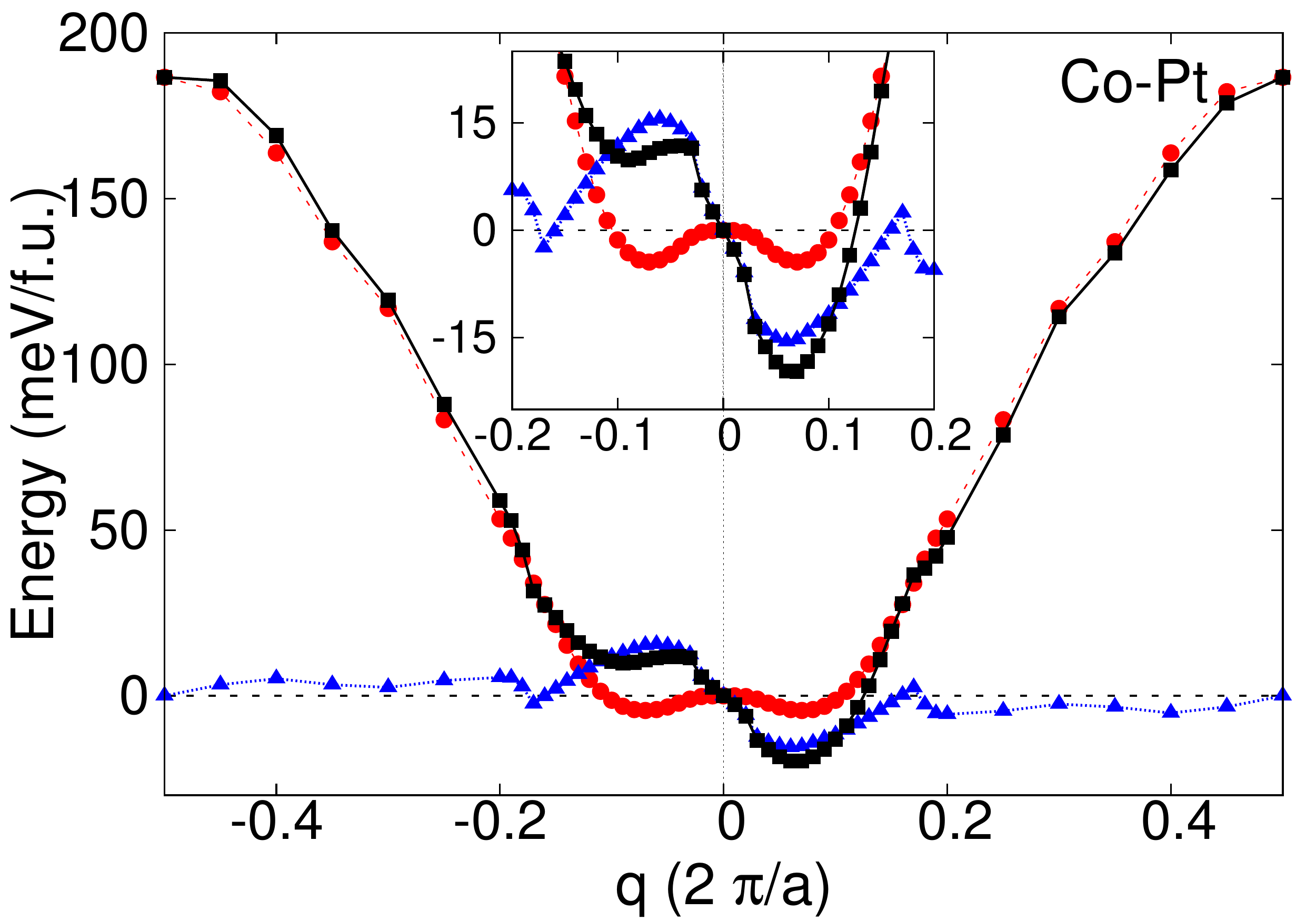}\\
       \vskip 0.1in
     \hspace{0.5cm} {\large  (c)} \hspace{7.2cm} {\large (f)}\\
    \includegraphics[angle=270,width=2.8in]{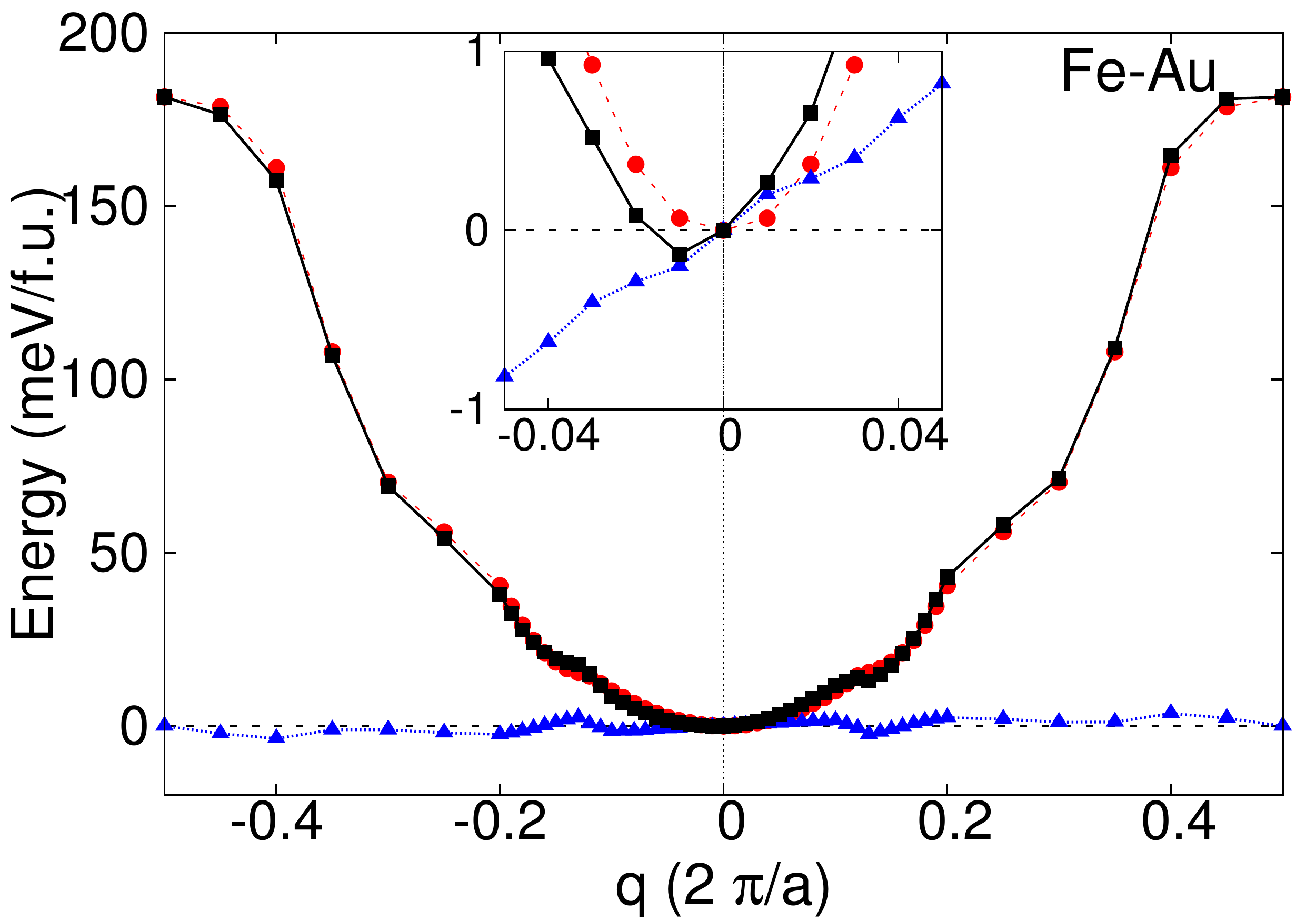}\hskip 0.3in \includegraphics[angle=270,width=2.8in]{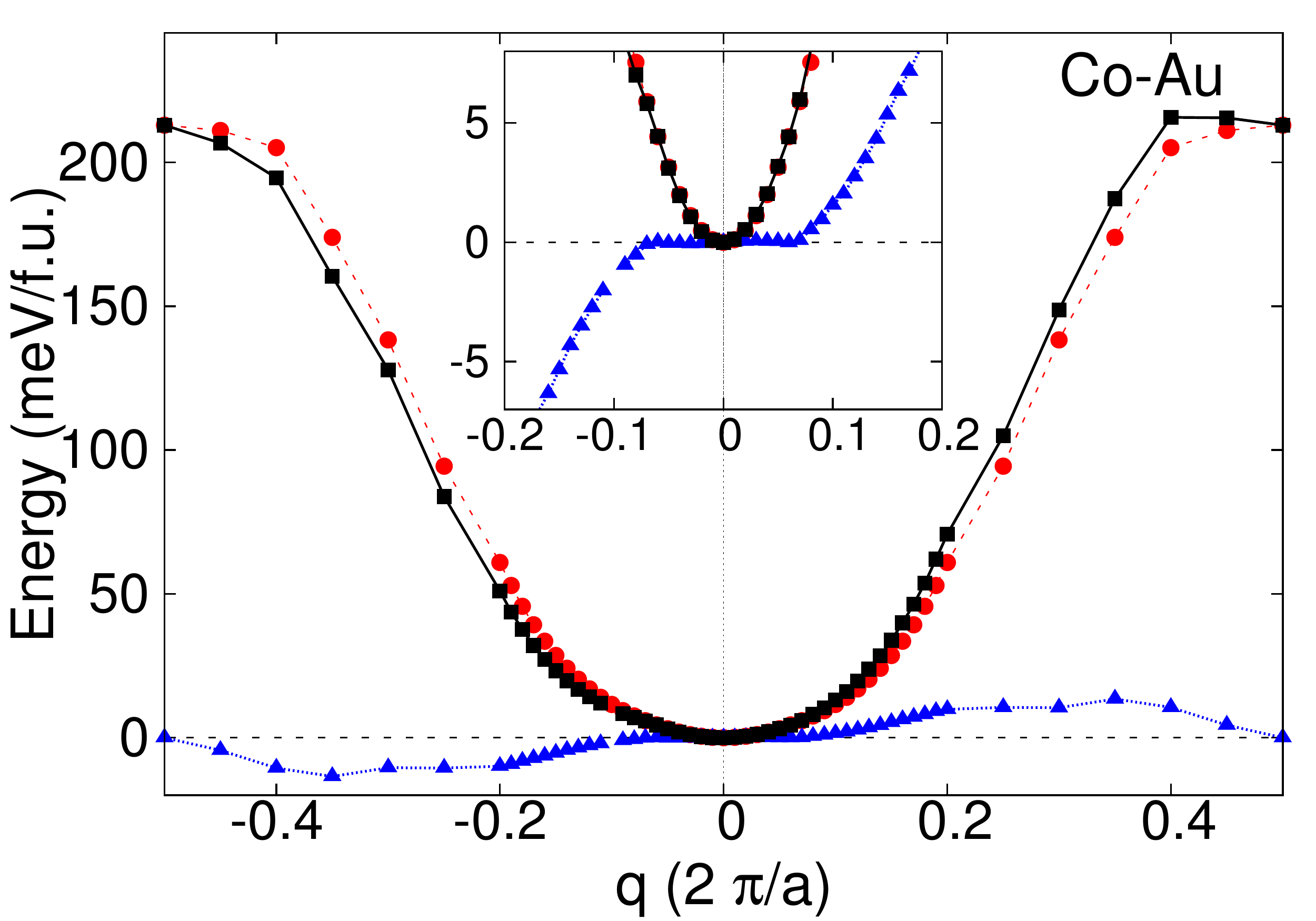}\\
  \caption{(Color online) The isotropic exchange-interaction energy $E_0(q)$ (red circles), the Dzyaloshinskii-Moriya interaction $E_{\rm DM}(q)$ (blue triangles) and the sum of both (black squares) for the various 3$d$-5$d$ chains as a function of the spin-spiral vector $q$ is displayed, calculated using LDA-VWN functional. The inset in each figure shows the magnified view in the vicinity of ferromagnetic state.}
    \label{DMI}
\end{figure*}

In order to investigate the DMI, we made use of the generalized Bloch theorem applied to the magnetic state of a flat homogeneous spin spiral and have included the SOI within the first-order perturbation theory as explained in Sec. \ref{SecMethDMI}. 
The calculated energy contribution due to the DMI, $E_{\rm DM}(q)$, is shown in Fig.~\ref{DMI} for all the 3$d$-5$d$ chains, once as plain values and once in addition to the exchange energy $E_0(q) $.  We find that for all wave vectors with $|q|\lesssim 0.08$ (recall all wave vectors are given in units of $2\pi/a$), the DMI energy is linear in the wave vector, $E_{\rm DM}(q)\approx Dq$, around the ferromagnetic state, $q=0$, and the sign of $D$, which determines the potential handedness of the magnetic structure changes sign from plus to minus and plus again when changing the $5d$ atom from Ir to Pt and then to Au. The $E_{\rm DM}$ vary on a scale of 5-15 meV and changes sign several times for one half of the Brillouin zone, {\it e.g.}\ for Fe-Pt at $q$-values of $0.25$ and $0.4$. Obviously,  $E_{\rm DM}(q)$ does not follow the simple $\sin q$ behavior for $0\leq |q|\leq 0.5$, but contains additional oscillations indicating that the DM vectors $\mathbf{D}_{ij}$ beyond the nearest neighbor interaction contribute significantly for larger wave vectors. 

Now we concentrate on the effect of DMI on the ground state. Therefore, we compare the energy minimum of $E(q) = E_0(q) + E_{\rm DM}(q)$ to the average MAE, where $E_0(q)$ is the isotropic spin-spiral dispersion energy (cf.\ insets in Fig.~\ref{DMI}).

\paragraph{Fe-5d chains:}
Let us recall the results from Sec.~\ref{chap_isotropic_exchange}: in the absence of SOI, Fe-Ir and Fe-Au chains are ferromagnetic, whereas the Fe-Pt chain shows a degenerate non-collinear ground state at $q=\pm0.03$. For the Fe-Ir chain, the DMI lowers the energy of right-handed spin spirals around the ferromagnetic state with an energy minimum of $E(q)$ at $q=-0.02$ (cf.\ Fig.~\ref{DMI}a). This energy minimum is 0.5 meV/f.u.\ lower than the ferromagnetic state, and thus DMI is too weak to compete against the average MAE of $1.2$ meV/f.u.\ (cf.\ Tab.~\ref{3d5d-MAE-DMI-t}). Similarly, in Fe-Au chains the DMI prefers right-handed spin spirals (cf.\ Fig.~\ref{DMI}c), but the energy gain of 0.1 meV/f.u.\ with respect to the ferromagnetic state is too small to compete against the average MAE (cf.\ Tab.~\ref{3d5d-MAE-DMI-t}). However, in Fe-Pt chains the strong DMI lifts the degeneracy in the spin-spiral ground state in favor of the left-handed spin-spiral with a significant energy gain of 7.1~meV/f.u.\ compared to the minimum of $E_{0}(q)$ (cf.\ Fig.~\ref{DMI}b). The energy minimum with respect to the ferromagnetic state of 8.5~meV/f.u.\ is an order of magnitude larger than the average MAE, leading to a left-rotating spin-spiral ground state with $q=+0.05$ corresponding to a wave length of $51$~\AA\ or $20$ lattice constants. 

\paragraph{Co-5d chains:}
In the absence of SOI, Co-Ir and Co-Au chains exhibit a ferromagnetic ground state (cf.\ Sec.~\ref{chap_isotropic_exchange}), and Co-Pt a degenerate non-collinear ground state. This picture does not change when including SOI. In Co-Ir and Co-Au chains, the DMI is too weak to compete against the average MAE (cf.\ Tab.~\ref{3d5d-MAE-DMI-t}). Interestingly, the $E_{\rm DM}$ vanishes for Co-Au in a relatively large region for $\lvert q \rvert<0.08$ (cf.\ Fig.~\ref{DMI}f). In contrast, the effect of DMI on the ground state in Co-Pt chains is strongest among the systems investigated in this paper (cf.\ Fig.~\ref{DMI}e) and lifts the degeneracy in favor of a left-handed spin spiral at $q=+0.07$ corresponding to a wave length of $36$ ~\AA\ or $14$ lattice constants. The large additional energy gain of 15.3 meV/f.u.\ leads to an energy minimum of $E(q)$ being 19.7 meV/f.u.\ lower than the ferromagnetic state (compared to an average MAE of only 0.1 meV/f.u.).

We estimate the inhomogeneity of the spin spirals in Fe-Pt and Co-Pt by extracting micromagnetic parameters $A$ and $D$ from fits to the energy dispersion $E_0(q)$ (for $\abs{q}<0.2$) and $E_\mathrm{DM}(q)$ (for $\abs{q}<0.05$), respectively. The DMI in those chains is so strong, that the inhomogeneity parameter $\kappa$ is rather tiny, $\kappa < 0.04$ for Fe-Pt and 0.004 for Co-Pt, {\it i.e.} the spirals are to a very good approximation homogeneous.

\begin{table}
\begin{center}
\resizebox{3.3in}{!}{
\begin{tabular} {c c c c c c } \hline \hline
  System & $\frac{1}{2} K_{1}$ & $\underset{q}{\rm min}\,E(q)$ & ~ $D(3d)$ ~ &~  $D(5d)$  ~ & $D$\\
 \hline
         &  (meV/f.u.) & (meV/f.u.)   & (meV\AA)  & (meV\AA) & (meV\AA) \\
  \hline \hline
  Fe-Ir & $1.2$  &  $-\phantom{0}0.5$  &  $\phantom{+}4.7$  &  $ \phantom{+0}26\phantom{.0} $  &  $ \phantom{+0}31\phantom{.0}$ \\
  Fe-Pt & $0.4$  &  $-\phantom{0}8.5$  &  $\phantom{+}0.7$  &  $-\phantom{0} 88\phantom{.0} $  &  $-\phantom{0}88\phantom{.0}$ \\
  Fe-Au & $0.9$  &  $-\phantom{0}0.1$  &  $\phantom{-}0.5$  &  $ \phantom{+00}6.3           $  &  $ \phantom{+00}6.8$ \\
  \hline
  Co-Ir & $7.1$  &  $-\phantom{0}4.7$  &            $-3.4$  &  $ \phantom{+0}20\phantom{.0} $  &  $ \phantom{+0}16\phantom{.0}$ \\
  Co-Pt & $0.1$  &  $-19.7$            &            $-8.8$  &  $-           124\phantom{.0} $  &  $-           132\phantom{.0}$ \\
  Co-Au & $0.7$  &  $\phantom{+0}0.0$  &            $-1.6$  &  $ \phantom{+00}2.5 $            &  $ \phantom{+00}0.8$  \\
  \hline
 \end{tabular}}
 \vskip 0.1in
 \caption{The comparison of the average magneto-crystalline anisotropy energy $K_{1}$ with the minimum of $E(q) = E_0(q) + E_{\rm DM}(q)$. The columns for  $D$, $D(3d)$ and $D(5d)$ represent the total strength of DMI of the system and decomposed for $3d$ atom and  $5d$ atom, respectively, in the vicinity of $q=0$.}
 \label{3d5d-MAE-DMI-t}
\end{center}
 \end{table}

\begin{figure*}
    {\large (a)}     \includegraphics[angle=270,width=5.4in]{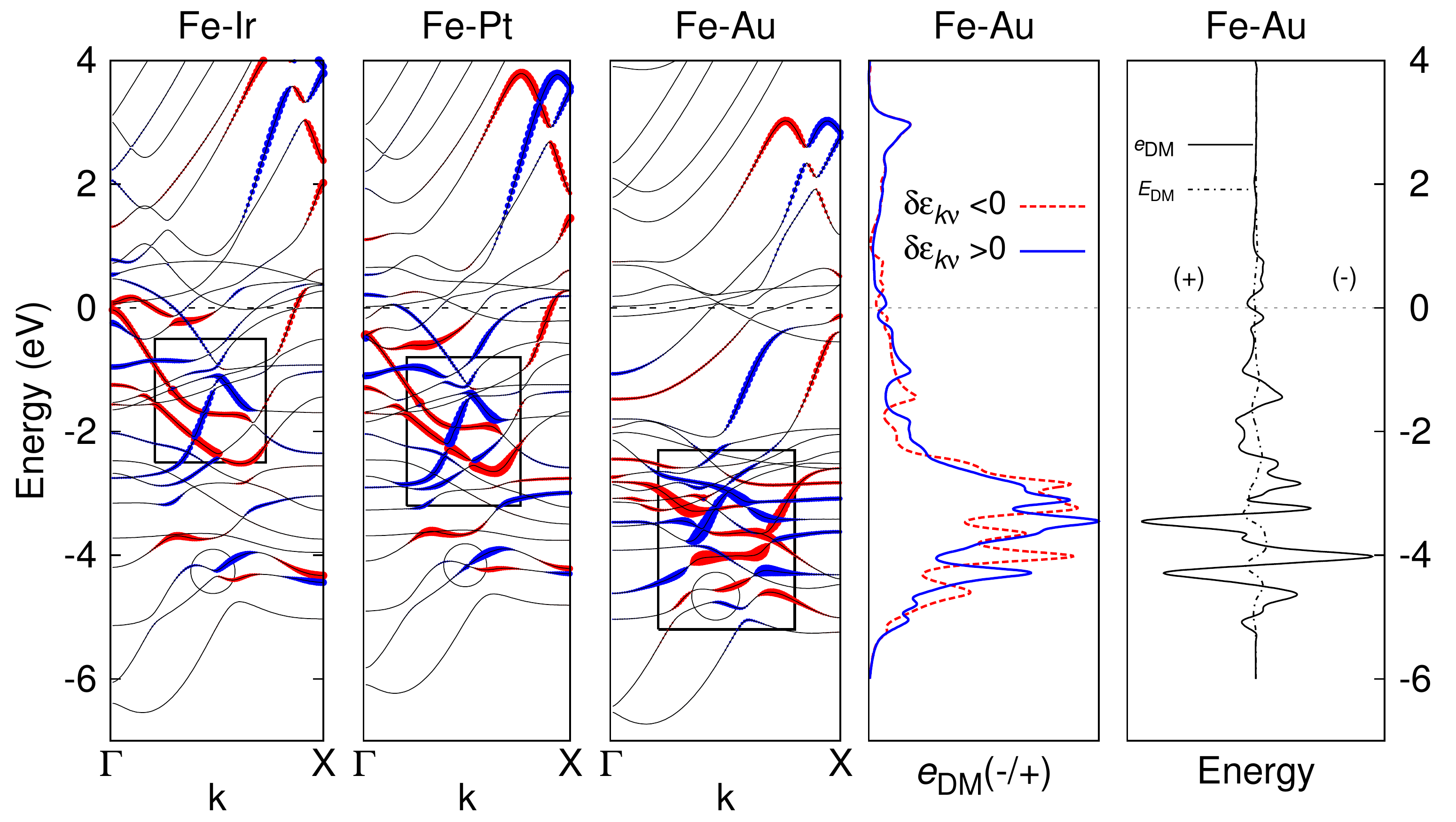} \\
    \vskip 0.0in
{\large (b)}    \includegraphics[angle=270,width=5.4in]{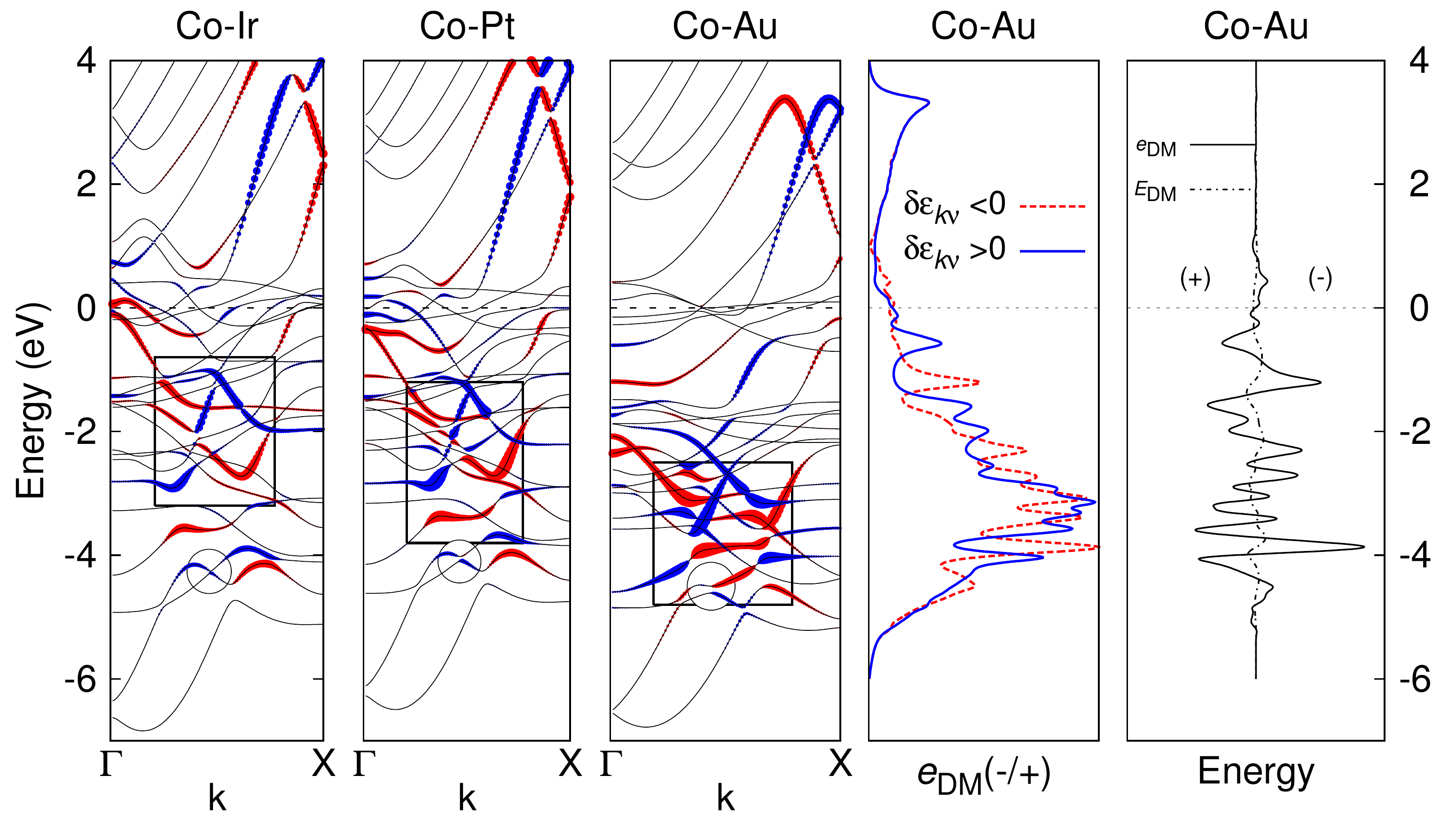}\\
    \vskip 0.0in
    \caption{(Color online) The relativistic electronic band structures, {\it i.e.}\ calculated including the SOI, of bi-atomic (a) Fe-5$d$ chains, and (b) Co-5$d$ chains, are shown for a spin-spiral state with a wave vector of $q=0.15$. The blue (red) dots indicate a positive (negative) shift of the spectral energies $\delta \epsilon_{k\nu}$, {\it i.e.}\ towards higher (lower) single particle energies, with respect to the scalar-relativistic eigenvalues, {\it i.e.}\ calculated without SOI. For visual clarity, the radius of the dot represents the actual difference in eV magnified 140 times. The highlighted rectangle shows the area in the band structure exhibiting the maximum SOI effect. The Fermi energy was chosen as origin of the energy scale. Panel 4 shows the energy resolved DMI contribution $e_{\rm DM}(\epsilon)$, separated into a curve originating from positive (blue) and negative (red) shifts only, for the $3d$-Au chains. In panel 5, the sum of the two and the effective energy-integrated DMI energy $E_{\rm DM}(\epsilon)$ is shown.}
    \label{SOC}
\end{figure*}

In order to investigate the effect of SOI  on the strength of DMI, we have decomposed the DMI into contributions from $3d$ and $5d$ transition-metal chain atoms, collected in Table~\ref{3d5d-MAE-DMI-t}.
We find interesting trends across the atomic species considered in our calculations: the contributions to $D$ for a specific atomic species is always of the same sign, \textit{e.g.}\ Fe atoms always yield a positive contribution to $D$ independent of the 5$d$ atom. The same holds for a specific 5$d$ atom.
Furthermore, we find that for the Ir and Pt chains with their large induced $5d$ magnetic moments and the spin-polarized $5d$ states, the Dzyaloshinskii-Moriya strength is solely determined by those $5d$ metals. The $5d$ atoms contribute to the effective DM vector, $D$, by about one order of magnitude more than the $3d$ atoms. This can be different for the Au chains. Au atoms exhibit a rather small spin-polarization of basically $s$ and $p$ electrons and their contribution to the $D$ vector can be of the same order as the one of the $3d$ metals, as our calculation shows. Also for the Co-Au chain the contributions of the Co and Au atoms to the DM vector are of similar size but opposite sign, and the total contribution cancels resulting in a $D$ vector with size close zero, at least on a size that is at the verge of the numerical resolution. 

It is worth noticing that the sign of  $D$ in the $3d$-Ir and $3d$-Pt zigzag chain follows exactly the sign found in respective $3d$ films. For Fe on Ir(111)~\cite{Heinze_11} the right-rotating $D$ leads to the nanoskyrmion structure in this system, while for Co/Pt(111)~\cite{Freimuth_14} a left handed $D$ was calculated and for  Co/Pt~\cite{Parkin_13} and FeCo/Pt~\cite{Beach_13} left-handed chiral domain walls were observed.

\begin{figure*}
    \centering
    \includegraphics[angle=270,width=6.5in]{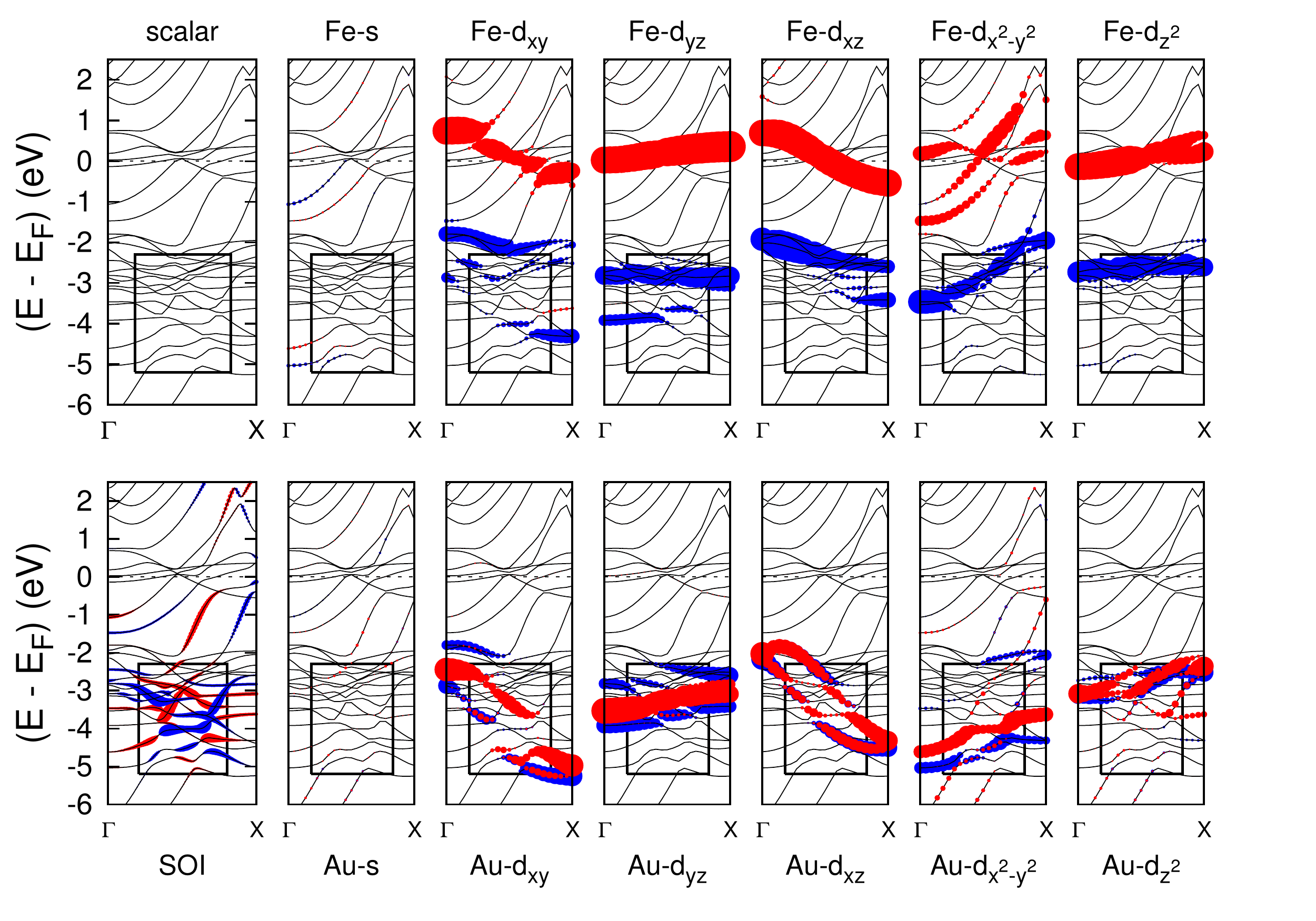}
     \caption{(Color online) Band structure without SOI (labeled as `\textit{scalar}') of the bi-atomic Fe-Au chain for $q=0.15$, and analysed in terms of the orbital character on the Fe and Au site. The Fermi energy, $E_F$, is chosen as origin of energy scale. In this figure, we have shown only $s$ and $d$-projected orbitals. The majority (minority) states are highlighted by blue (red) dots. The radius of the dots at each point in the band structure $(k\nu)$ is proportional to the respective orbital character. Additionally, also the relativistic band structure including the shift of the spectral energies due to SOI is shown in the lower left panel. The highlighted rectangle shows the area in the band structure where maximum SOI effect is observed.}
    \label{fat}
\end{figure*}

The SOI affects different parts of the Brillouin zone, different bands and even different parts of a single band differently. To provide an understanding in how the electronic structure of a chain is effected by the SOI we present in the first three panels of Fig.~\ref{SOC}(a) and \ref{SOC}(b) the one-dimensional relativistic band structure along the high-symmetry line $\Gamma$--X for the bi-atomic Fe and Co zigzag chains, respectively, for the same spiral magnetic state with a  spin-spiral vector chosen to be $q=0.15$.  The effect of SOI results in a change of the energy dispersion of the Bloch states. The energy $\epsilon$ of the states $(k\nu)$ is shifted with respect to the scalar-relativistic (SR) treatment, {\it i.e.}\ neglecting the spin-orbit interaction, of electronic structure by an amount $\delta \epsilon_{k\nu}= \epsilon_{k\nu}^\mathrm{SOI} - \epsilon_{k\nu}^\mathrm{SR}$. These shifts are highlighted by dots, whose size is proportional to $|\delta \epsilon_{k\nu}|$. A shift to higher (lower) binding energy, $\delta \epsilon_{k\nu}<0$ ($>0$), is indicated by red (blue) dots. At the first glance, we see that the topologies of the six band structures are very similar. They are determined by exchange-split $3d$ states and the $5d$ states of Ir, Pt, and Au. They differ in the band width and the  position of those states with respect to the Fermi energy. The Au $d$-states are all below the Fermi energy. Pt and Ir have one and two electrons, respectively, less and their $d$ states at the edge of the $5d$ valence band move through the Fermi energy. For Fe and Co,  the majority $3d$ states are all below the Fermi energy and the minority $d$-states cross the Fermi energy.  For more details we refer to the discussion of Fig.~\ref{fat}. 

Considering now the shifts, $\delta \epsilon_{k\nu}$, we find colored dots with significantly larger radii as compared to the bands in the rest of the Brillouin zone  basically located in the bands related to the  $5d$ states. Therefore, their energy position with respect to the Fermi energy depends only on the $5d$ atom of the zigzag chain and not on the $3d$ one. We have highlighted this region of the band structure by enclosing it within a rectangle. This region of large shifts moves up towards the Fermi energy when changing from Au to Ir atoms, just as the $5d$ states move upwards.  The actual size of the shifts depends on the hybridization between the $3d$ majority states and the $5d$ states, and this hybridization becomes smaller if the $5d$  states move up when changing the Au atom by Pt or Ir, while the $3d$ states stay at energy where they are. This energy is the same for both Co and Fe majority states and therefore the size and position of shifts depend only on the $5d$ metal atom of the chain.

The fourth panel of Fig.~\ref{SOC}(a) and \ref{SOC}(b) shows the energy resolved DMI contribution for positive and negative shifts, $\delta \epsilon_{k\nu}$, when integrated over the Brillouin zone, $e_\textrm{DM}(\epsilon,q)=\sum_\nu\int \delta \epsilon_{k\nu}\,\delta(\epsilon-\epsilon_{k\nu})\,dk$, and smoothened by a Lorentzian function $\frac{1}{\pi}\frac{\Gamma / 2}{(\delta\epsilon-\delta\epsilon_{k\nu})^{2}+(\Gamma / 2)^{2}}$ with a full width at half maximum of 0.2~eV, and plotted as function of the binding energy. We have calculated the DMI distribution for all the chains, however, in Fig.~\ref{SOC}(a) and \ref{SOC}(b) only the results for Fe-Au and Co-Au chains are shown. The last panel of Fig.~\ref{SOC} shows the effective energy integrated DMI energy, $E_\textrm{DM}(\epsilon,q)=\int^{\epsilon} e_\textrm{DM}(\epsilon^\prime,q)\,d\epsilon^\prime$, calculated from the positive and negative SOI shifts of the band structure. We observe that for both chains, Fe-Au and Co-Au, the energy resolved DMI has the largest contribution at a binding energy of around 3.5 ~eV. From what is said above, there is no surprise that the maximum is around the same energy for both chains, as the maximum depends basically on the $5d$ atom. In detail $e_\textrm{DM}(\epsilon)$ are slightly different for both chains due to the difference in the hybrization of the Fe and Co $3d$ electrons with the Au $5d$ ones. Since all $5d$ states of Au are below the Fermi energy the integral of the energy resolved DMI contribution up to the Fermi energy for positive and negative shifts are nearly the same and  $E_\textrm{DM}(E_\textrm{F},q)$ is very small. 
The energy resolved DMI contribution for positive and negative shifts are nearly the same and can in first approximation be thought to be rigidly shifted by about 0.6~eV. Due to this finite shift of $e_\textrm{DM}(\epsilon)$ between positively and negatively shifted states, $E_\textrm{DM}(\epsilon)$ oscillates as function of the band filling. We observe a rapidly oscillating function of large Dzyaloshinskii-Moriya energies of oscillating signs, particularly in the center of the Au $5d$ bands. For example, the first significant peak we find at about $-1$~eV and then large peaks at around $-4$~eV.  When the Au atom is replaced by a $5d$ metal atom with less $d$ electrons, $E_\textrm{DM}(\epsilon)$ moves relative to $E_\textrm{F}$. Assuming a rigid band model where the $5d$ band does not change upon changing $5d$ metal we can adopt the $5d$ electron number such that the Fermi energy is placed in one of those peaks. This happens actually for the Pt and Ir chains. The Fermi energy moves into the regime of the large peaks which explains the large contribution of Pt and Ir to the DM vector as discussed in Table~\ref{3d5d-MAE-DMI-t} and explains the sign change of $D$ between Pt and Ir chains moving the Fermi energy by about 0.4~eV. 

In this sense the  $E_\textrm{DM}(E)$ allows a design of the strength and the sign of the Dzyaloshinskii-Moriya interaction by selecting the number of $5d$ electrons such that the Fermi energy $E_\textrm{F}$ is in the right ball-park of the peak. To realize a chain with the optimal number of $5d$ electrons one may require an alloyed  zigzag chain, where the  $5d$ atom site is occupied randomly by different $5d$ atoms with a particular concentration. Then, additional {\it ab initio} calculations might be necessary for a fine-optimization of the composition, overcoming the assumptions made in the rigid band model.

In the discussion of Fig.~\ref{SOC}, we have identified regions in the band structure, where SOI effects are large. To get a better understanding of the underlying microscopic mechanism, we have performed a {site-,} orbital- and spin-resolved analysis of the scalar-relativistic band structure. In a spin-spiral calculation, the up- and down-states are calculated with respect to the local spin-quantization axis in each muffin tin sphere. The resulting contributions are shown in Fig.~\ref{fat} for the Fe-Au chain for $q=0.15$. The energy bands showing largest SOI effect in the band structure are mainly the Au-$d_{xy}$, $d_{xz}$ and $d_{yz}$ states hybridizing with the Fe majority-states.
It can be inferred that, the effective DMI contribution obtained from the positive and negative shifts is maximal where the SOI effect as well as the orbital hybridization is maximal.

\section{$d$-orbital tight-binding model of the Dzyaloshinskii-Moriya interaction in isosceles trimers}  \label{sec-tb}

\subsection{The model}
\begin{figure}
    \centering
    \includegraphics[width=2.6in]{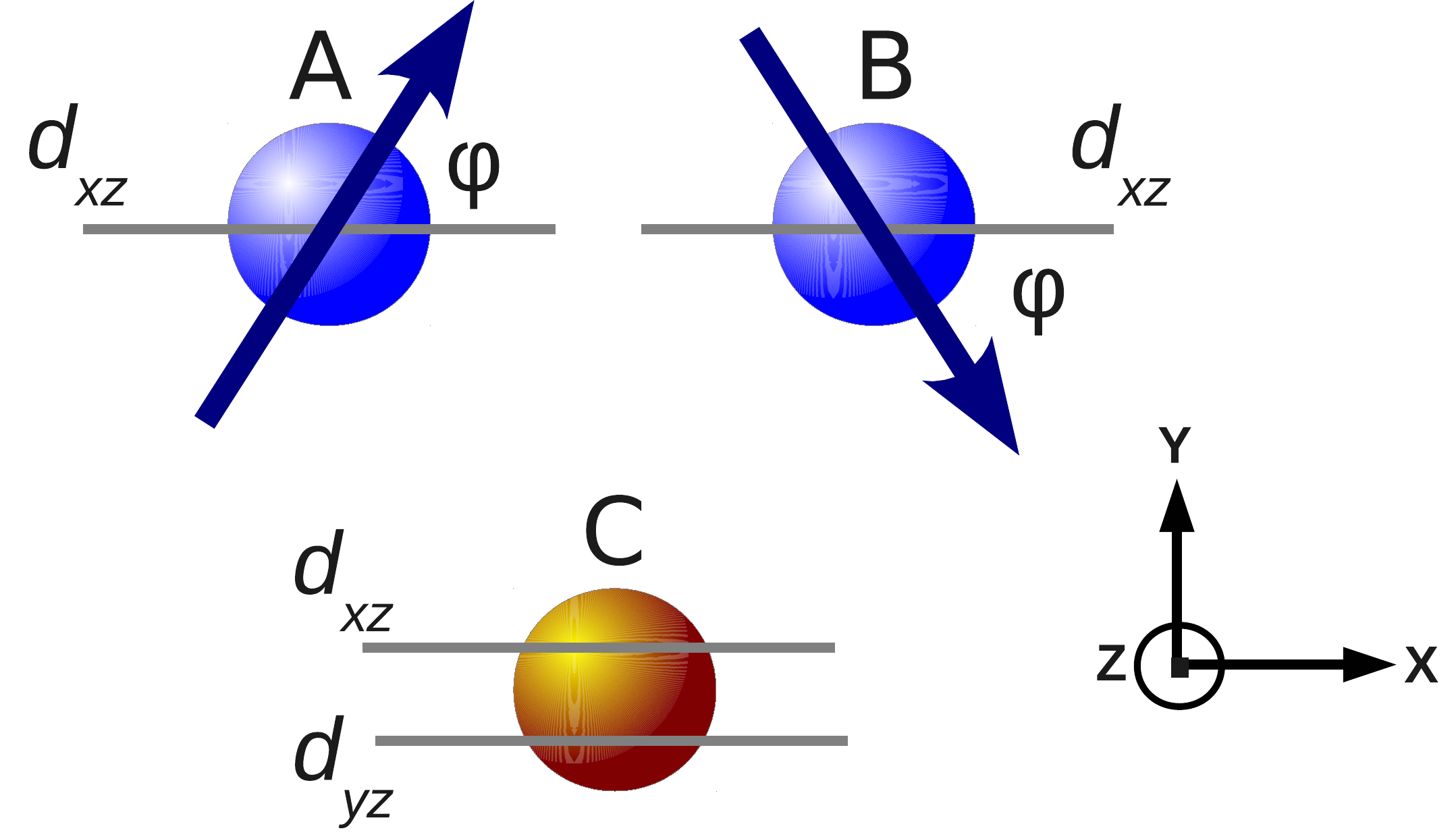}
    \caption{(Color online) The structure of the trimer. It consists of two magnetic sites A and B represented by blue spheres with a $d_{xz}$ orbital and one non-magnetic site C represented by a brown sphere with the $d_{xz}$ and $d_{yz}$ orbital. The magnetic moments of the magnetic sites are canted by an angle $\varphi$ with respect to the $x$-axis, in which the angle for the magnetic moment of the first/second site is defined in a mathematically positive/negative rotation sense (see also Eq.~\eqref{mag_moments}). We denote this magnetic structure as right-rotating with a chirality vector $\bm{c} = \bm{S_A} \times \bm{S_B} = - \bm{e_z}$.}    
    \label{trimer}
\end{figure}

The minimal model exhibiting a non-vanishing Dzyaloshinskii-Moriya interaction that can be associated with the bi-atomic zig-zag chains discussed above is an isosceles trimer made of two identical 3$d$-metal sites carrying no spin-orbit interaction (SOI) and one non-magnetic 5$d$-metal site having a large SOI. In this context, the word non-magnetic stands here for zero intrinsic on-site exchange splitting. However, hybridization with the magnetic sites will lead to a small induced spin-polarization at the non-magnetic site and thus to a small magnetic moment after the calculation. The magnetic sites will be denoted as A and B and the non-magnetic site as C, henceforth. Without loss of generality the trimer is arranged within the  $x$-$y$ plane (Fig.~\ref{trimer}).

The model is based on a tight-binding description restricted to the two energetically degenerate $d_{xz}$- and $d_{yz}$-orbitals for the non-magnetic site and only the $d_{xz}$-orbital, with the same on-site energy at each magnetic site. According to our analysis in section \ref{DMI_abinitio}, these orbitals are those yielding the main contributions to the Dzyaloshinskii-Moriya interaction for this specific geometry. The following 8$\times$8 Hamiltonian in representation of the basis set $(d_{xz}^\mathrm{A}, d_{xz}^\mathrm{B}, d_{xz}^\mathrm{C}, d_{yz}^\mathrm{C})$ with the superscripts denoting the site index and with the $x$-axis chosen as global spin-quantization axis reflects this 8-state model: 
\begin{widetext}
\begin{equation}
\Hamiltonian = \left(
\begin{array}{c c c c | c c c c}
E_\mathrm{A}-Im/2 \cos \varphi  & 0 & \phantom{+}t_1 & \phantom{+}t_2 & \mathrm{i} Im/2 \sin \varphi & 0 & 0 & 0 \\
0 & E_\mathrm{A}-Im/2 \cos \varphi & \phantom{+}t_1 & -t_2 & 0 & -\mathrm{i} Im/2 \sin \varphi & 0 & 0 \\
\phantom{+}t_1 & \phantom{+}t_1 &  E_\mathrm{C} & 0 & 0 & 0 & 0 & \mathrm{i} \xi/2 \\
\phantom{+}t_2 & -t_2 &  0 & E_\mathrm{C} & 0 & 0 & -\mathrm{i} \xi/2 & 0 \\
\cline{1-8} -\mathrm{i} Im/2 \sin \varphi & 0 & 0 & 0 & E_\mathrm{A}+Im/2 \cos \varphi & 0 & \phantom{+}t_1 & \phantom{+}t_2  \\
0 & \mathrm{i} Im/2 \sin \varphi  & 0 & 0 & 0 & E_\mathrm{A}+Im/2 \cos \varphi & \phantom{+}t_1 & -t_2 \\
0 & 0 & 0 & \mathrm{i} \xi/2 & \phantom{+}t_1 & \phantom{+}t_1 &  E_\mathrm{C} & 0 \\
0 & 0 & -\mathrm{i} \xi/2 & 0 & \phantom{+}t_2 & -t_2 &  0 & E_\mathrm{C} \\
\end{array} 
\right),
\label{8_band_model}
\end{equation}
\end{widetext}
where $E_\mathrm{A}$($=E_\mathrm{B}$) and $E_\mathrm{C}$ are the on-site energies, $t_1$ and $t_2$ are the hopping parameters between atoms A, B with atom C, $\varphi$ is the angle of the magnetic moments relative to the quantization axis, $I$ is the Stoner parameter of the magnetic sites and $m$ the corresponding magnetic moment and $\xi$ is the spin-orbit strength. The separation into the 4$\times$4 sub-blocks highlights the $\uparrow\uparrow$, $\uparrow\downarrow$, $\downarrow\uparrow$ and $\downarrow\downarrow$ spin-blocks of $\Hamiltonian$. In the following, the model Hamiltonian \eqref{8_band_model} will be discussed in detail.

The Hamiltonian $\Hamiltonian$ comprises three contributions, 
\begin{equation}
 \Hamiltonian = \Hamiltonian_{\mathrm{0}} + \Hamiltonian_{\mathrm{mag}} + \Hamiltonian_{\mathrm{SO}},  \label{TB_Hamiltonian} \end{equation}
where $\Hamiltonian_{0}$ contains the spin-independent hopping elements and the on-site energies of the system, $\Hamiltonian_{\mathrm{mag}}$ incorporates magnetism and $\Hamiltonian_{\mathrm{SO}}$ introduces the spin-orbit interaction. 

The hopping matrix elements of $\Hamiltonian_{0}$, $t_1$ and $t_2$ in Eq.~\eqref{8_band_model}, describe the electron transition between the $d_{xz}$ orbital at the magnetic sites and the $d_{xz}$ or  $d_{yz}$ orbitals, respectively, on the non-magnetic site. We employed the  Slater-Koster parametrization\cite{SlaterKoster}  requiring two Slater-Koster parameters $V_{dd\pi}$ and $V_{dd\delta}$\footnote{The third Slater-Koster parameter $V_{dd\sigma}$ is not needed since the trimer lies within the $x$-$y$ plane.} that determine the matrix elements as:
\begin{eqnarray}
t_1 & = & \hat{R}_x^2 V_{dd\pi} + \hat{R}_y^2 V_{dd\delta}, \\
t_2 & = & \hat{R}_x \hat{R}_y (V_{dd\pi} - V_{dd\delta}),
\end{eqnarray}
where $\hat{R}_x$ and $\hat{R}_y$ are the direction cosines of the bonding vector between the sites involved in the hopping. This follows from the Slater-Koster transformations\cite{SlaterKoster} for our specific geometry and choice of orbitals. Since direct hopping between the magnetic sites is not necessary to obtain a non-vanishing Dzyaloshinskii-Moriya interaction, this minimal model is restricted to $t_1$ and $t_2$ only. Obviously $t_2\propto  \hat{R}_y$ and thus $t_2$ scales with the structure inversion asymmetry of our trimer model, {\it i.e.}\ $t_2$ becomes zero if the trimer changes from a triangular to a chain geometry. The on-site energies are denoted as $E_\mathrm{A}$ for both magnetic sites and $E_\mathrm{C}$ for the non-magnetic site. Note, to simplify our model they depend only on the site and not on the type of orbital.

To investigate the Dzyaloshinskii-Moriya interaction (DMI) magnetism is incorporated within the Stoner model,\cite{Stoner_model1,Stoner_model2} extended to the description of non-collinear magnetic systems:\cite{TB_noco,TB_noco2}
\begin{equation} \Hamiltonian_{\mathrm{mag}}= -\frac{1}{2} I\,  {\bm {m}} {\bm \sigma}, \end{equation}
where ${\bm \sigma}$ is the Pauli vector. The exchange splitting of the electronic structure depends on the Stoner parameter $I$ and the magnetic moment ${\bm m_{i}}$ of the magnetic sites only, whereas no intrinsic exchange splitting is assumed at the non-magnetic site. Due to symmetry, only the rotation of the magnetic moments within the $x$-$y$ plane is of interest in the determination of the DMI, as discussed in Sec.~\ref{SecMethDMI}. Therefore, the site-dependent magnetic moment is 
\begin{equation} {\bm m_{i}} = m_i(\cos{\varphi} \cdot {\bm e_{x}} \pm \sin{\varphi} \cdot {\bm e_{y}}) \lbl{mag_moments} \end{equation} 
with the plus-sign for the site A and the minus-sign for site B, respectively. $\varphi$ is the angle of the magnetic moment within the $x$-$y$-plane with respect to the $x$-axis (see Fig.~\ref{trimer}).

Since the DMI is the consequence of the spin-orbit interaction (SOI), SOI has to be implemented into the tight-binding model by expressing the term $  {\bm \sigma} {\bm L}$ within the atomic orbital representation. By introducing a SOI parameter $\xi$ for the non-magnetic site and taking into account that the interaction is on-site, the SOI matrix reads
\begin{equation} [\Hamiltonian_{\mathrm{SO}}]_{\mu\nu}^{\sigma\sigma'}=\frac{1}{2} \xi \langle {\mu \sigma} \lvert {\bm \sigma} {\bm L} \lvert {\nu \sigma'} \rangle,  \end{equation}
where $\mu$, $\nu$ indicate the orbitals and $\sigma$, $\sigma'$ are the indices of the spin. In this model the only non-zero matrix element of $\Hamiltonian_{\mathrm{SO}}$ is the spin-flip element between the $d_{xz}$ and $d_{yz}$ orbital at the non-magnetic site, \mbox{$\langle {d_{xz}^\mathrm{C}\uparrow} \lvert {\bm \sigma} {\bm L}  \lvert {d_{yz}^\mathrm{C}\downarrow} \rangle = \mathrm{i}$}. 

Typically the spin-orbit interaction (SOI) is a small contribution to the entire Hamiltonian, hence it is common to calculate the SOI energy contribution and therefore also the Dzyaloshinskii-Moriya interaction within first-order perturbation theory.\cite{Heide} The simple 8-state model can be easily solved by diagonalizing the Hamiltonian $\Hamiltonian$ of Eq.~\eqref{8_band_model}, which contains SOI, however to allow for a qualitative comparison with the previously presented zig-zag chain results, SOI is treated within first-order perturbation theory. That means, the Hamiltonian $\Hamiltonian_{0}+\Hamiltonian_{\mathrm{mag}}$ is diagonalized and the eigenvalues $\varepsilon_{n}$ and eigenvectors $\lvert n \rangle$ are used to determine the contributions
\begin{equation} \delta \varepsilon_{n}=\langle n \lvert \Hamiltonian_{\mathrm{SO}} \lvert n \rangle , \label{1st_order}  \end{equation}
similar to Eq.~\eqref{delta_e_abintio}.
The Dzyaloshinskii-Moriya interaction (DMI) energy can be calculated via the expression
\begin{equation} 
E_{\mathrm{DMI}}= \sum_{n} \delta \varepsilon_{n} \cdot f(\varepsilon_{n}), \label{E_DMI} 
\end{equation}  
where $f(\varepsilon_{n})$ displays the Fermi-Dirac occupation function. This equation corresponds to Eq.~\eqref{EsumSOC} in the case of a finite system. Since the only non-zero matrix element in $\Hamiltonian_{\mathrm{SO}}$ is the transition between $d_{xz}^{\uparrow} \to d_{yz}^{\downarrow}$ and vice versa, it is the only transition which is at the end responsible for $\delta \varepsilon_{n}$ and $E_{\mathrm{DMI}}$.

\subsection{Results}
For the calculations following parameters have been used, which are chosen to be reasonable values for 3$d$ and $5d$ transition-metal systems: the Slater-Koster parameters $V_{dd\pi}=0.8$~eV and $V_{dd\delta}=-0.07$~eV lead to the hopping parameters $t_1=0.148$~eV and $t_2=-0.377$~eV. The on-site energies of the magnetic sites are $E_\mathrm{A}=E_\mathrm{B}=0$~eV and $E_\mathrm{C}=1 $~eV for the non-magnetic site. A Stoner parameter of $I=0.96$~eV and magnetic moments of $m_A=m_B=1.2$~$\mu_B$ lead to an exchange splitting of $1.152$~eV. The spin-orbit interaction parameter $\xi$ of the non-magnetic site is $0.6$~eV. The system is occupied by 6 electrons.

\begin{figure} [!ht]
\centering
(a) $\varphi=0^\circ$:
\includegraphics[width=0.43\textwidth]{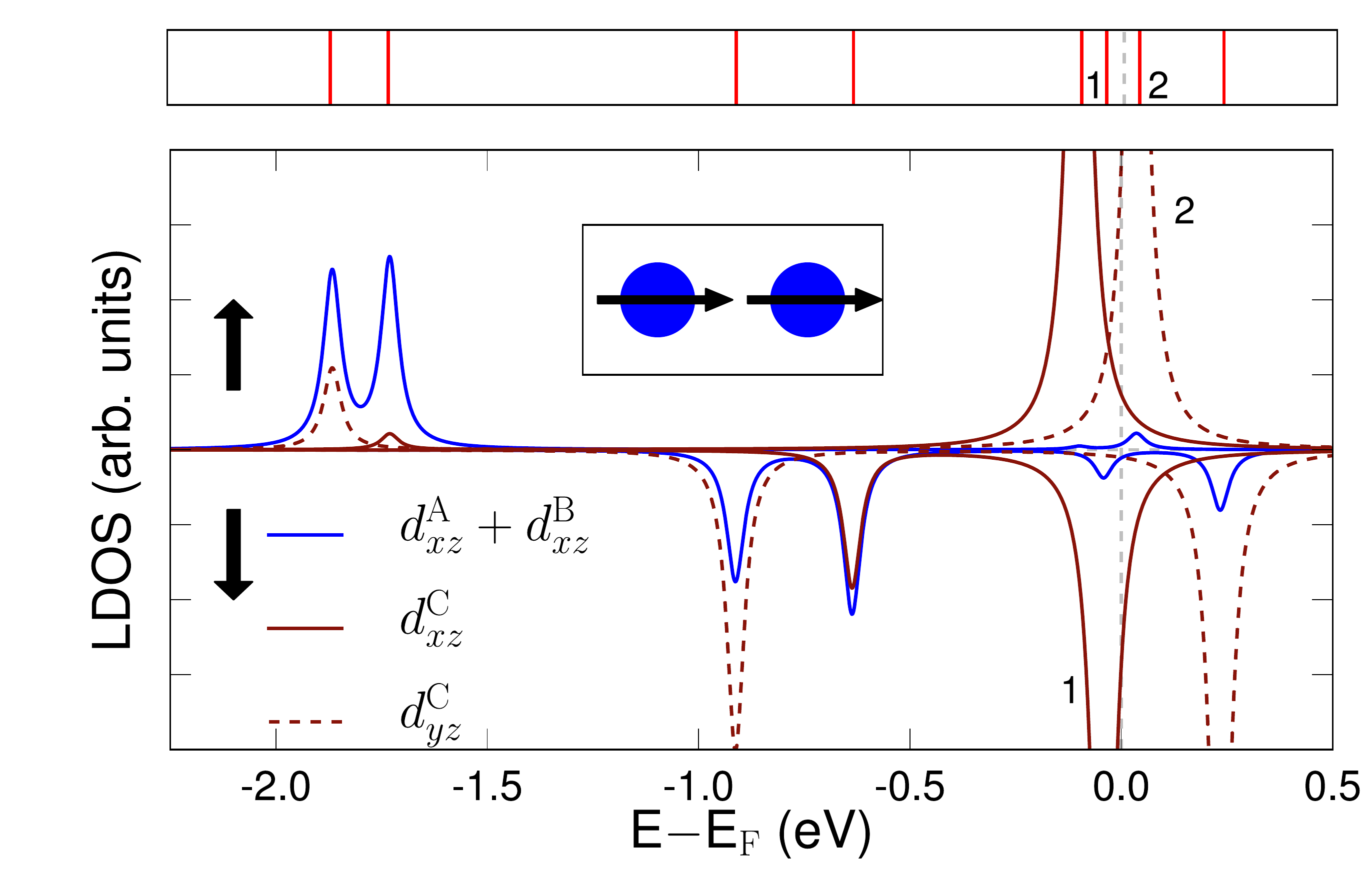}
\hspace{2cm}(b) $\varphi=45^\circ$:
\includegraphics[width=0.43\textwidth]{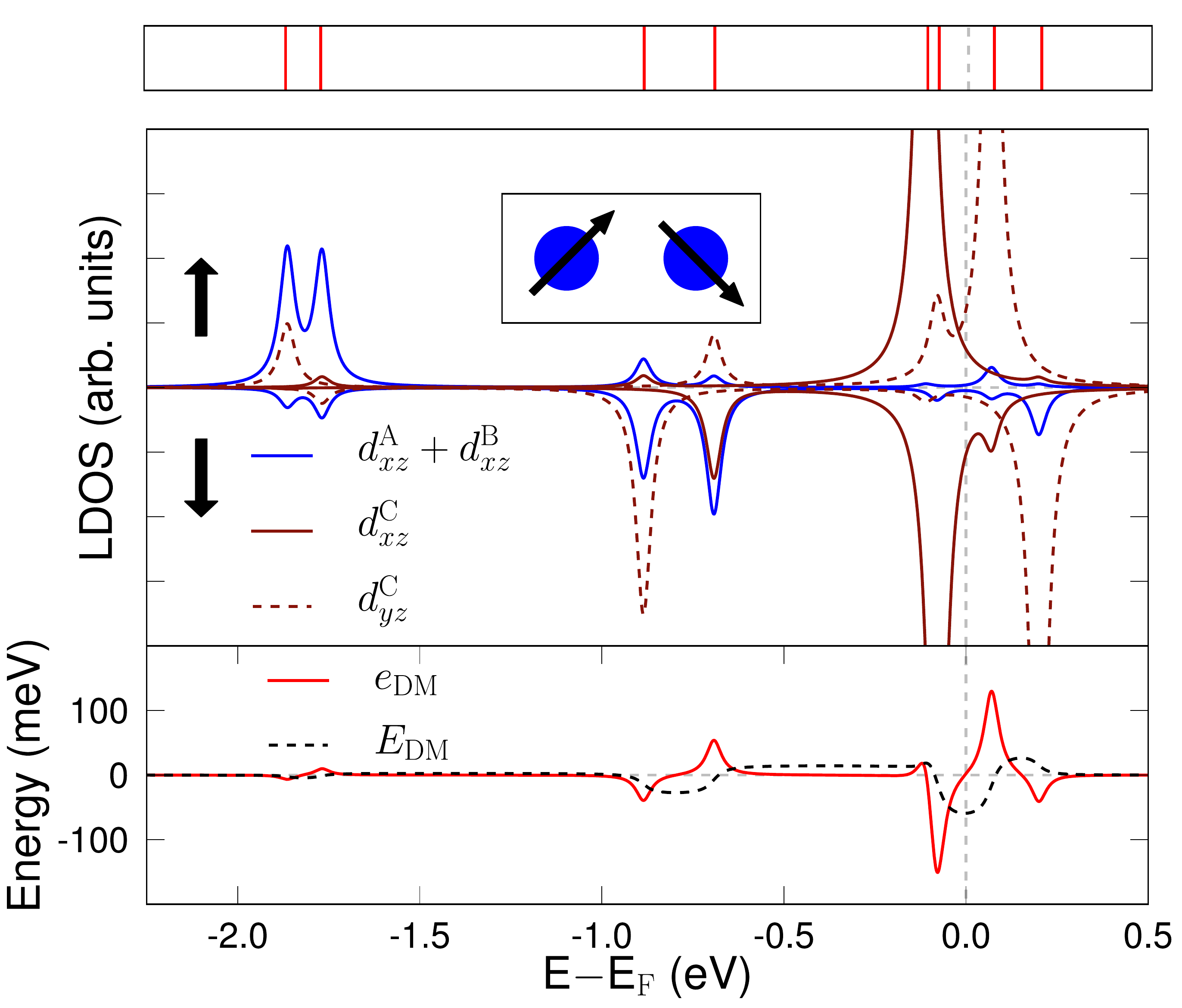}
\caption{(Color online) The site-, orbital- and spin-resolved density of states (DOS) of the trimer is displayed for two magnetic configurations (a) $\varphi=0^\circ$ and (b) $\varphi=45^\circ$. In each case, the 8 eigenenergies are displayed in an upper panel above the respective DOS. The upper (lower) panel of the spin-resolved DOS describes the majority (minority) states with respect to the spin-quantization axis chosen along the $x$-direction. The blue curve displays the DOS of the two magnetic sites in the $d_{xz}$-orbital (denoted as $d_{xz}^\mathrm{A}+d_{xz}^\mathrm{B}$) and the brown solid, dashed curve represents the DOS of the $d_{xz}$- and $d_{yz}$-orbital of the non-magnetic site (denoted as $d_{xz}^\mathrm{C}$ and  $d_{yz}^\mathrm{C}$). Note that the DOS of the non-magnetic site has been enhanced by a factor of 5. In the case of $\varphi=45^\circ$, $e_\textrm{DM}$ (red solid curve) and the integrated value $E_\textrm{DM}$ (black dashed curve) is displayed in the lower panel. More information on these two quantities and details about the two eigenenergies 1 and 2 and the corresponding DOS can be found in the text.}
\label{noco_case}
\end{figure}

First the role of magnetism for the Dzyaloshinskii-Moriya interaction (DMI) is discussed by comparing the density of states (DOS) of the ferromagnetic case ($\varphi=0$) with the maximally canted case of $\varphi=45^\circ$. The sign of $\varphi$ as defined in Fig.~\ref{trimer} is chosen such that the chiral magnetic structure is stable, {\it i.e.}\ the DMI energy is negative. Both results are displayed in Fig.~\ref{noco_case} and the analysis is conducted by comparing the site-, orbital- and spin-resolved DOS of the unperturbed system $\Hamiltonian_{0}+\Hamiltonian_{\mathrm{mag}}$ to get more insight into $e_\textrm{DM}$, which is the sum of all contributions $\delta \varepsilon_n$ due to the spin-orbit interaction broadened by Lorentzian functions (see also Sec.~\ref{DMI_abinitio}). For both, the DOS and $e_\textrm{DM}$ a broadening of $25$~meV with full width at half maximum was used. In addition the DMI energy $E_\textrm{DM}(E)$ is plotted, which is the integrated value of $e_\textrm{DM}$ up to an energy $E$. So $E_\textrm{DM}(E_{\rm F})$ corresponds to the definition of $E_\mathrm{DMI}$ of Eq.~\eqref{E_DMI}.

The DOS of Fig.~\ref{noco_case}(a) is easily understood in terms of our 8$\times$8 model \eqref{8_band_model}. The majority and minority channel consists of 4 states each. The energy distribution of the 4 states is a result of the hybridization between the $d_{xz}$ states at the 3$d$-metal sites and the $d_{xz}$ and $d_{yz}$ states at the 5$d$-metal site, with the bonding states at low energies and the antibonding states at energies around the Fermi energy. The energy splitting among the states results from the different hybridization between the $d_{xz}$-$d_{xz}$ and $d_{xz}$-$d_{yz}$ orbitals. The majority and minority states of the 3$d$-metal sites are shifted by an exchange splitting $Im$. Since the minority states are closer in energy to the states of the 5$d$-metal site, there the hybridization is larger. This hybridization leads also to a small exchange splitting of the states at the 5$d$-metal site.
   
If the magnetic moments of the magnetic sites are ferromagnetically aligned as in Fig.~\ref{noco_case}(a), no DMI can be observed,\footnote{For the antiferromagnetic alignment the asymmetric exchange vanishes due to a two-fold degeneracy of each eigenenergy.} and $e_\textrm{DM}$ vanishes, since an eigenstate has either pure $d_{xz}$- or $d_{yz}$-character of the non-magnetic site but not both, and Eq.~\eqref{1st_order} is zero. Due to the Lorentzian broadening, the eigenenergies 1 and 2 around the Fermi energy in Fig.~\ref{noco_case}(a) seem to contribute largely to the DMI. However, eigenenergy 1 exhibits only $d_{xz}^\downarrow$- and eigenenergy 2 only $d_{yz}^\uparrow$-character. Hence, their eigenfunctions cannot contribute to DMI. In contrast, the case of $\varphi=45^\circ$ of Fig.~\ref{noco_case}(b) shows that the non-collinearity of the magnetic sites is crucial to obtain a non-vanishing DMI. The $d_{xz}$-orbitals of the magnetic sites hybridize with the orbitals of the non-magnetic site differently for different spin-channels and induce a spin-polarization. As a consequence the eigenstates obtain both $d_{xz}$- and $d_{yz}$-character of different spin leading to a non-zero $e_\textrm{DM}$.

The quantity $e_\textrm{DM}$ shows an interesting characteristic behavior. Each peak-like contribution comes along with an energetically slightly shifted contribution of opposite sign. This leads to sign changes in $E_\textrm{DM}$ if the Fermi energy is in the middle of such a feature. This explains the sensitivity of the magnitude and the sign of the asymmetric exchange depending on the substrate as presented in Sec.~\ref{DMI_abinitio}. Since $E_\textrm{DM}(E_{\rm F})$ corresponds to the DMI energy $E_\mathrm{DMI}$, an electron filling of 6 electrons leads to non-vanishing $E_\mathrm{DMI}$ in Fig.~\ref{noco_case}(b), whereas $E_\mathrm{DMI}$ vanishes for the maximum occupation number of 8 electrons in our finite system.

Beside the non-collinearity, the breaking of the inversion-symmetry is also crucial for the appearance of the DMI. Contrary to the non-inversion-symmetric trimer as displayed in Fig.~\ref{trimer}, the inversion-symmetric trimer\footnote{In inversion-symmetric trimer the non-magnetic and magnetic sites are located on one line.} exhibits no DMI (not shown). Here, no hybridization between the $d_{yz}$-orbital of the non-magnetic site and the $d_{xz}$-orbital of the magnetic sites occurs, since $t_2=0$. This again shows that the hybridization between the orbitals of the magnetic sites and the orbitals of the non-magnetic site is crucial, which can be also observed in the {\it ab initio} results of several zig-zag chains as presented in Fig.~\ref{fat}.

\begin{figure}[t!]
\centering
(a) $E_\mathrm{C} - E_\mathrm{A} = 3$~eV:
\includegraphics[width=0.43\textwidth]{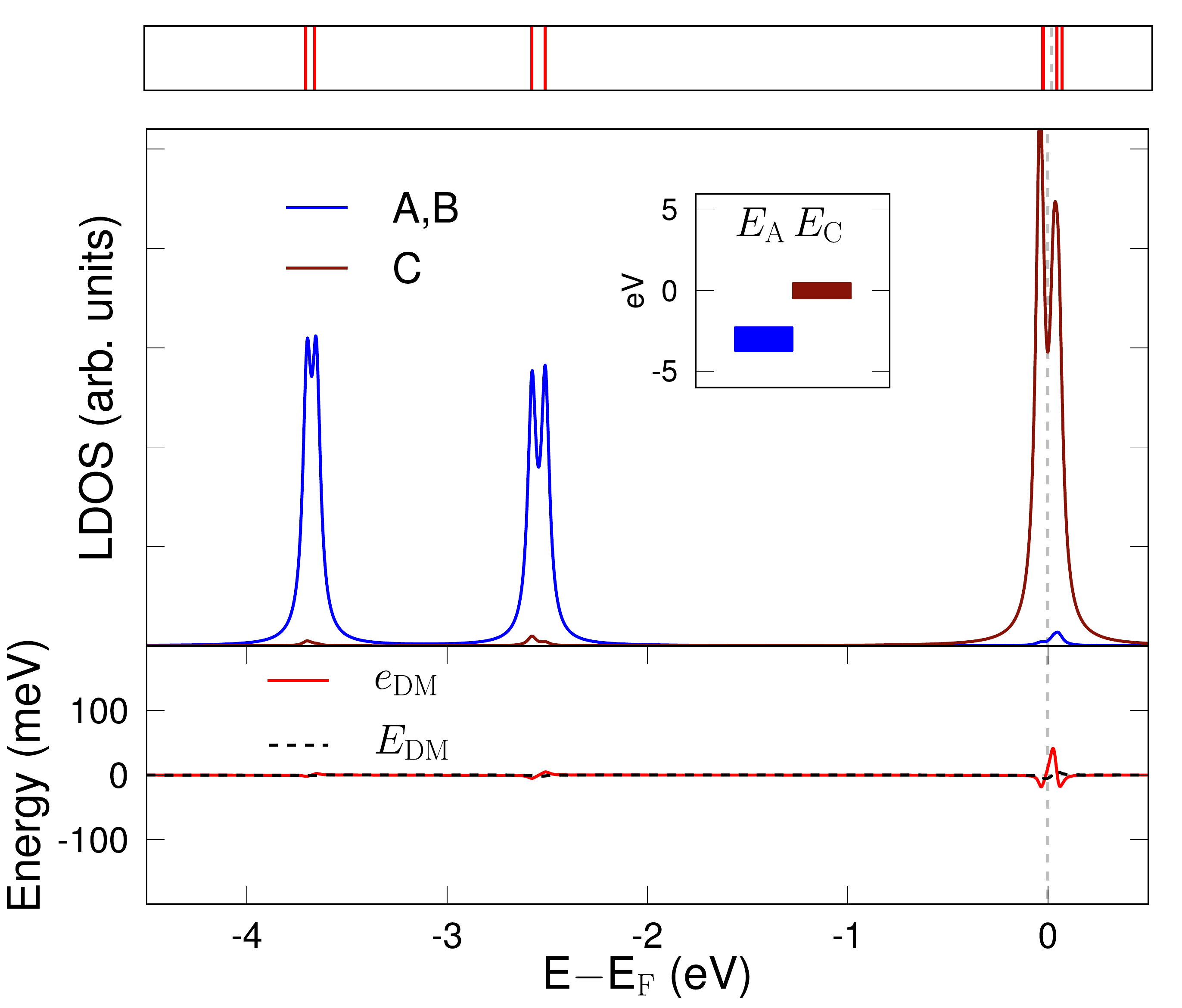}  
\hspace{2cm}(b) $E_\mathrm{C} - E_\mathrm{A} = 0.5$~eV: 
\includegraphics[width=0.43\textwidth]{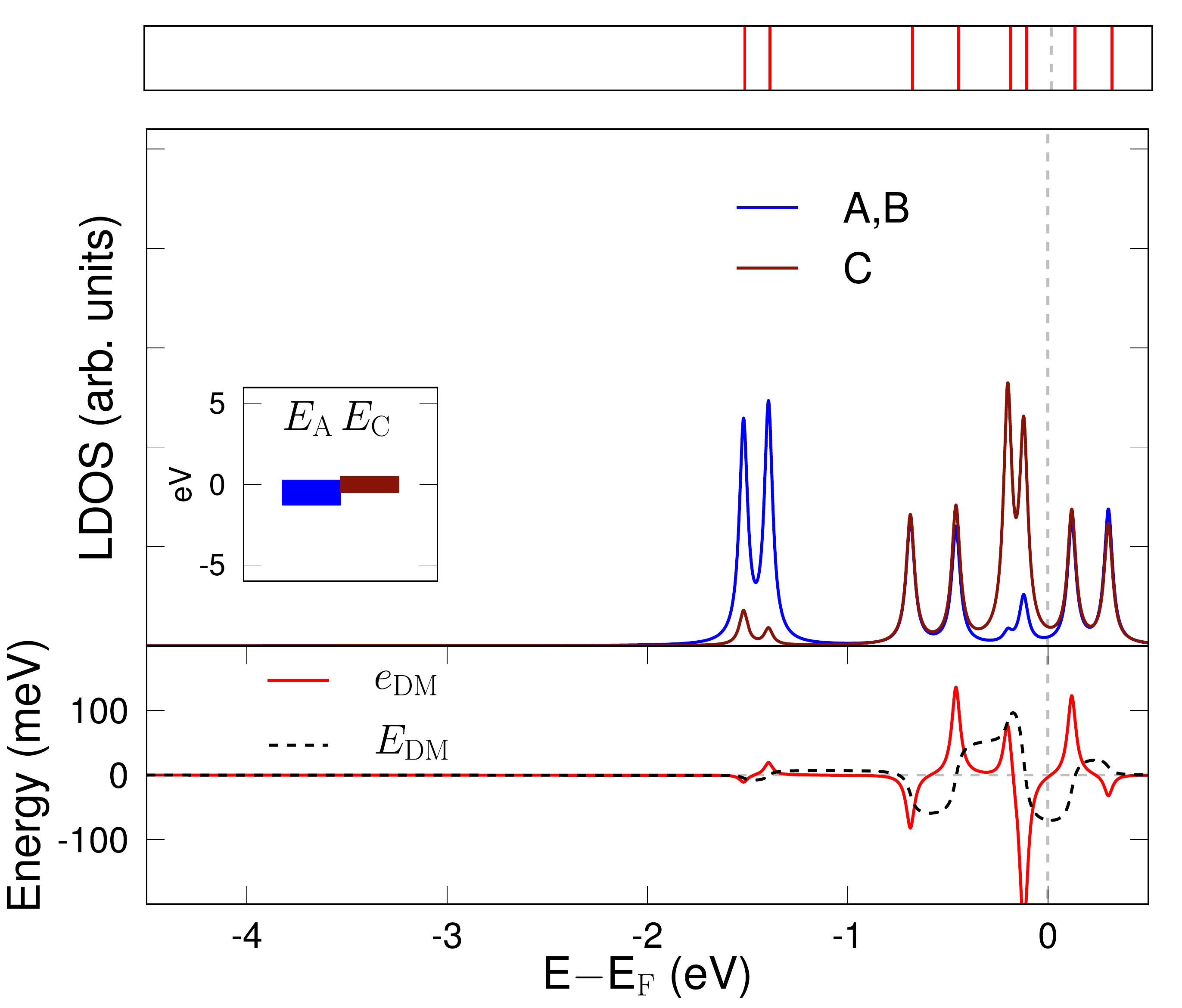}
\caption{(Color online) The site-resolved but not spin-resolved density of states (DOS) of the trimer within our tight-binding model is shown for two different on-site energy differences: (a) $E_\mathrm{C} - E_\mathrm{A} = 3$~eV and (b)  $E_\mathrm{C} - E_\mathrm{A} = 0.5$~eV. The maximally canted case with $\varphi=45^\circ$ resulting in a negative DMI energy has been considered. The 8 eigenenergies of the system are displayed in the upper panel for each case. The site-resolved DOS is displayed above with the blue curve showing the DOS of the magnetic sites (denoted as A,B) and the brown curve the DOS of the non-magnetic site (denoted as C). The energy-resolved first-order SOI contribution $e_\textrm{DM}$ (red solid curve) and the integrated quantity $E_\textrm{DM}$ (black dashed curve) are shown in the lower panels. The inset displays the on-site energy difference between the non-magnetic site and the magnetic sites (brown: C, blue: A and B). A reasonable bandwidth of about 1-2~eV is indicated by the width of the blue and brown boxes.} 
\label{compare_onsite_cases}
\end{figure}

It is interesting to take a look at the DMI as function of the difference between the on-site energies $E_\mathrm{C} - E_\mathrm{A}$, which controls the degree of hybridization between the magnetic sites and non-magnetic site. The results are summarized in Fig.~\ref{compare_onsite_cases}. The DMI becomes larger in magnitude for smaller on-site energy differences, as can be seen by comparing the magnitude of $e_\textrm{DM}$ in the lower panel of Fig.~\ref{compare_onsite_cases}(a) and Fig.~\ref{compare_onsite_cases}(b). In the case (a), the orbitals of the non-magnetic site and the magnetic sites hardly overlap due to a large on-site energy difference of 3~eV. Hence, the DMI is much smaller compared to the case (b) with an on-site energy difference of just 0.5~eV for which the orbitals are strongly hybridizing with each other. Fig.~\ref{compare_onsite_cases} demonstrates again the sensitivity of the magnitude and sign of $E_\textrm{DMI}$ on the details of hybridization between the strongly magnetic 3$d$ and the heavy 5$d$ transition-metal atom. In a simplified picture the main difference between {\it e.g.}\ an Fe-Pt and a Co-Ir zig-zag chain can be seen in the differences in the on-site energies, since the total number of electrons is the same for both systems. The sensitivity of the DMI on the substrate, as presented in Table \ref{3d5d-MAE-DMI-t} can be verified in this simple model. 

All results of this section have been calculated by adding spin-orbit interaction (SOI) within first-order perturbation theory. Although the results obtained by diagonalizing the full Hamiltonian of Eq.~\eqref{8_band_model} might be numerically different as compared to the results obtained within perturbation theory, the same conclusions on the behavior of the DMI can be drawn. The difference between these treatments is expected to be much smaller for the zig-zag chains of Sec.~\ref{DMI_abinitio}, since there the energy shifts $\delta E$ within first-order perturbation theory are an order of magnitude smaller as compared to the about $0.1$~eV changes in this simple model.

\subsection{Spin-orbit interaction and non-collinear magnetism within first-order perturbation theory}
The Dzyaloshinskii-Moriya interaction (DMI) turns out to be a very delicate quantity. Even for the simplified model Hamiltonian \eqref{8_band_model} it is illuminating to investigate the  sign and the magnitude of the DMI  treating the spin-orbit interaction and the non-collinear alignment of the quantization axis at the different atoms A and B within first-order perturbation theory. 
We start from the unperturbed system, which is the ferromagnetically ($\varphi=0$) aligned trimer with magnetic moments pointing along the $x$-axis, and without spin-orbit interaction.  Under these conditions the Hamiltonian (\ref{8_band_model}) block-diagonalizes into two $4\times 4$ Hamiltonians for majority and minority states, {\it i.e.}\ $\Hamiltonian_0+\Hamiltonian_\mathrm{mag}(\varphi=0) = {\cal H}^{\uparrow\uparrow}\oplus {\cal H}^{\downarrow\downarrow}$. The eigenstates of this magnetic system are denoted as $\vert n\rangle=\vert n^\sigma \rangle\oplus {\bm 0}$, for the first four eigenstates, $n=1,\dots 4$, which  correspond to the majority states ($\sigma=\uparrow$ or $\sigma=1$), and $\vert n\rangle={\bm 0}\oplus\vert n^\sigma \rangle $ for the second set of four eigenstates, $n=5,\dots 8$, which  correspond to the four  minority states ($\sigma=\downarrow$ or $\sigma=-1$). The four site- and orbital-dependent components of $\vert n^\sigma \rangle $ are denoted by $n^\sigma_{\mu^i}$, where $\mu^i$ is the atomic orbital $\mu$ at site $i$.
The matrix elements of the unperturbed Hamiltonian, $\Hamiltonian_0+\Hamiltonian_\mathrm{mag}(\varphi=0)$, are real numbers and  thus also the eigenstates $\lvert n^\sigma \rangle$ can be chosen to be real. The analytical  solution of the eigenenergies $\varepsilon_{n}^{\sigma}$ and eigenvectors $\vert n^\sigma \rangle$  read:
\begin{eqnarray}
\label{eq:8eigenvectors}
\varepsilon_{n}^{\sigma} & = & \frac{1}{2}\left(E_\mathrm{A}-\sigma\frac{1}{2} Im+E_\mathrm{C}\right) \nonumber \\
 & \, &+ \tau \sqrt{\frac{1}{4}\left(E_\mathrm{A}-\sigma\frac{1}{2} Im-E_\mathrm{C}\right)^2+2t_i^2} , \\
 \vert n^\sigma \rangle & = & \frac{1}{\sqrt{2+\frac{4t_i^2}{(\varepsilon_{\tau,i}^{\sigma}-E_\mathrm{C})^2}}} \nonumber \\
& \, & \times \left\{ \begin{array}{ll}
 \left( 1, \phantom{+}1, \frac{2t_1}{\varepsilon_{\tau,1}^{\sigma}-E_\mathrm{C}},0  \right)^{\mathrm{T}} & \textrm{for  $i=1$}\\
\left( 1, -1, 0, \frac{2t_2}{\varepsilon_{\tau,2}^{\sigma}-E_\mathrm{C}}  \right)^{\mathrm{T}} & \textrm{for  $i=2$}
\end{array} \right. . 
\end{eqnarray}
The 8 eigenvalues and eigenvectors characterized by $n= (\tau,i,\sigma)$ are a result of the hybridization of majority ($\sigma=1$) or minority ($\sigma=-1$) states at site A and B with either the  $d_{xz}$ ($i=1$) or $d_{yz}$ orbital ($i=2$) at site C, corresponding to the hopping parameter $t_i=t_1$ or $t_2$ that lead to bonding ($\tau=-1$) or antibonding ($\tau=+1$) states. Notice, the eigenvectors have either  $d_{xz}$ or $d_{yz}$ character at site C, but not both.  Although we deal here with a discrete eigenvalue spectrum in a more abstract sense we interpret the first term of the eigenenergies  as the center and the second term as half of a bandwidth $W_n^{\sigma}$, which is the energy difference between the corresponding bonding and antibonding state. These quantities will be used later when displaying the energy correction $\delta \varepsilon_n$.
 
First, let us evaluate the perturbed eigenvector $\lvert n \rangle$ under the perturbation of a small exchange field $\bm{B}=\gamma  \bm{e_y}$ along the $y$-direction, leading to a slightly canted magnetic configuration with angle $\varphi$ on magnetic site A. The parameter $\gamma$ is directly related to the angle $\varphi$ and therefore it is connected to the degree of non-collinearity. Note that this canted configuration is different from that depicted in Fig.~\ref{trimer} in that we keep the spin on site B along the $x$-axis to simplify the notation. The corresponding perturbation of the Hamiltonian in spin-space reads:
\begin{equation}
\Delta V  =  -\Theta_{\mathrm{A}}^\dagger \bm{\sigma}  \bm{B}\, \Theta_\mathrm{A}  
          =  -\gamma \sigma_y \delta_\mathrm{AA},
\end{equation}   
where $\Theta_\mathrm{A}$ is a step function which is zero outside atom A, and $\delta_\mathrm{AA}$ is a projection onto orbitals localized on site A, with the first part of this equality being basis-independent, and the right side being in the representation of the localized atomic orbitals. Henceforth, all equations will be given in both a basis-independent form as well as the atomic orbital representation. In the representation of the $x$-axis as spin-quantization axis, $\sigma_y$ contains only spin-flip elements, leading to changes to the unperturbed eigenstate $\lvert n^\sigma \rangle$ that are of purely opposite spin-contribution $\sigma'$. Hence, the following equation for $\vert n \rangle$ in first-order perturbation theory with $\sigma \neq \sigma'$ is obtained:
\begin{eqnarray}
\lvert n \rangle  & = &  \lvert n^\sigma \rangle + \sum_{\substack{n'\, (\neq n)\\ (\sigma'\neq\sigma)}} \frac{\langle n'^{\sigma'} \lvert \Delta V \lvert n^\sigma \rangle}{\varepsilon_{n}^{\sigma} - \varepsilon_{n'}^{\sigma'}} \lvert n'^{\sigma'} \rangle \\
& = & \lvert n^\sigma \rangle - \mathrm{i} \gamma \sum_{\substack{n'\, (\neq n)\\ (\sigma'\neq\sigma)}} \frac{\delta S_{n'^{\sigma'} n^\sigma}}{\varepsilon_{n}^{\sigma} - \varepsilon_{n'}^{\sigma'}} \lvert n'^{\sigma'} \rangle ,
\label{perturbed_state}
\end{eqnarray}
where $\varepsilon_{n}^{\sigma}$ and $\varepsilon_{n'}^{\sigma'}$ are the corresponding eigenvalues to the unperturbed eigenstates $\lvert n^\sigma \rangle$ and $\lvert n'^{\sigma'} \rangle$ and the real quantity 
\begin{eqnarray}
\delta S_{n'^{\sigma'} n^\sigma} & = &  -\mathrm{i} \langle n'^{\sigma'} \lvert   \Theta_{\mathrm{A}}^\dagger  \sigma_{y}  \Theta_\mathrm{A}   \lvert n^\sigma \rangle \\
& = & -\mathrm{i} \langle n'^{\sigma'} \lvert \sigma_y \lvert n^\sigma \rangle_{\rm A} =  \sigma n'^{\sigma'}_{d_{xz}^\mathrm{A}} n^\sigma_{d_{xz}^\mathrm{A}} \label{delta_S},
\end{eqnarray}
where by $\langle n'^{\sigma'} \lvert \sigma_y \lvert n^\sigma \rangle_{\rm A}$ we introduce a short-hand notation for the evaluation of the term $\langle n'^{\sigma'} \lvert \sigma_y \lvert n^\sigma \rangle$ at site A. The prefactor $\sigma$ is 1 ($-1$) depending whether $n^\sigma$ belongs to the  majority (minority) spin channel.

Next, we evaluate within first-order perturbation theory the correction to the energy of state $|n\rangle$ due to the spin-orbit interaction by substituting the perturbed state $\lvert n \rangle$ of Eq.~\eqref{perturbed_state} into Eq.~\eqref{1st_order}. Neglecting higher-order terms in $\varphi$ and taking into account that $\langle n^\sigma \lvert \Hamiltonian_{\mathrm{SO}} \lvert n^\sigma \rangle = 0$ leads to the following expression for the energy shift:
\begin{eqnarray}
\delta \varepsilon_{n} & = & 2\gamma \sum_{\substack{n'\, (\neq n)\\ (\sigma'\neq\sigma)}} \frac{\delta S_{n'^{\sigma'} n^\sigma}}{\varepsilon_{n}^{\sigma} - \varepsilon_{n'}^{\sigma'}}  \operatorname{Im} \langle n^\sigma \lvert \Hamiltonian_\mathrm{SO} \lvert n'^{\sigma'} \rangle \\
& = & \gamma \xi \sum_{\substack{n'\, (\neq n)\\ (\sigma'\neq\sigma)}} \frac{\delta L_{n^\sigma n'^{\sigma'}} \delta S_{n'^{\sigma'} n^\sigma}} {\varepsilon_{n}^{\sigma} - \varepsilon_{n'}^{\sigma'}} \label{deltaEn_pert},
\end{eqnarray}
with
\begin{eqnarray}
\delta L_{n^\sigma n'^{\sigma'}}  & = & \operatorname{Im} \langle n^\sigma \lvert \Theta{_\mathrm{C}}^\dagger \left( L_y \sigma_y + L_z \sigma_z \right)  \Theta_\mathrm{C} \lvert n'^{\sigma'} \rangle \\ 
& = & \frac{1}{2} \operatorname{Im} \langle n^\sigma \lvert L_{\mp} \sigma_{\pm} \lvert n'^{\sigma'} \rangle_{\rm C} \label{Lmp_ref} \\
& = & n'^{\sigma'}_{d_{yz}^\mathrm{C}} n^\sigma_{d_{xz}^\mathrm{C}} - n'^{\sigma'}_{d_{xz}^\mathrm{C}} n^\sigma_{d_{yz}^\mathrm{C}} ,
\end{eqnarray}
where in Eq.~\eqref{Lmp_ref} $L_\mp$ are the angular moment ladder operators and $\sigma_\pm$ the spin ladder operators with $L_{-}\sigma_{+}$ corresponding to the case $\sigma = \; \uparrow$ and $L_{+}\sigma_{-}$ to $\sigma = \; \downarrow$, respectively. For our particular situation the $L_y$ operator in the subspace of considered local orbitals vanishes, which allows to write the contribution to the energy shift due to SOI as:
\begin{eqnarray}
\delta \varepsilon_{n}  & = & -\gamma \xi \sum_{\substack{n'\, (\neq n)\\ (\sigma'\neq\sigma)}} 
\frac{\mathrm{i} \langle n'^{\sigma'} \lvert \sigma_y \lvert  n^{\sigma}  \rangle_{\rm A}\langle  n^{\sigma}  \lvert  L_z \sigma_z   \lvert  n'^{\sigma'}  \rangle_{\rm C}}{\varepsilon_{n}^{\sigma} - \varepsilon_{n'}^{\sigma'}} \nonumber \\ 
& = &- \gamma \xi  t_1 t_2 (-1)^{i}\sigma\!\! \sum_{\substack{n'\, (\neq n) \\ (i'\neq i)}} \frac{\tau\,\delta_{\sigma,-\sigma'}\tau'}{W_{n}^{\sigma}  (\varepsilon_{n}^{\sigma} - \varepsilon_{n'}^{\sigma'})W_{n'}^{\sigma'}}  . \label{YM}
\end{eqnarray}
The sum runs over the 8 states $n'=(\tau', i',\sigma')$, but has only non-zero summands if the  orbital-type $i'$ on site C and the spin direction $\sigma'$ of state $n'$ are different from the state $n=(\tau, i,\sigma)$. $\sigma=1(-1)$  stands for electrons  of state  $n^{\sigma}$ taken from the majority (minority) spin channel. According to Eq.~(\ref{eq:8eigenvectors}),  $i=1(2)$ corresponds to the state with a $d_{xz}$ ($d_{yz}$) orbital component. The product $\tau\tau'=1(-1)$ if the bonding character labelled by $\tau$ of both states is the same (different).
The quantities $W_n^{\sigma}$ and $W_{n'}^{\sigma'}$ play the role of the bandwidths, since they are the energy differences between the corresponding bonding and antibonding states of $\lvert n^{\sigma} \rangle$ and $\lvert n'^{\sigma'} \rangle$.

We remark, that in general the energy shifts due to spin rotation away from collinear configuration at the  different sites A and B are independent and should be added up in order to comprise the total shift $\delta\varepsilon_n$. For example, upon a simultaneous $\varphi$ and $-\varphi$ rotation of spins on A and B sites, respectively,  as shown in Fig.~\ref{trimer}, the $\delta\varepsilon_n$ from Eq.~(\ref{YM}) corresponding to a staggered B-field on both sites should be simply multiplied by a factor two, owing to symmetry. We confirmed that Eq.~\eqref{YM} reproduces well the energy shifts $\delta\varepsilon_n$ obtained by diagonalization of $H_0+H_{\mathrm{mag}}$ and including then the SOI through first-order perturbation theory as presented in Fig.~\ref{noco_case}. The error for a canting angle of $\varphi=45^\circ$ is about 10\%. 

The DMI energy, $E_\textrm{DMI}$~\eqref{E_DMI},  is then obtained by the summation over all the energy shifts $\delta \varepsilon_{n}$ of states $n$ that are occupied. By this, the sum over $n'$ in Eq.~\eqref{YM} changes to a double sum over $n$ and $n'$ and all those combinations of states $n$ and $n'$ in sum~\eqref{YM} cancel out identically if occupied both.  Thus, at the end only those combinations of states contribute to the DMI energy for which the  initial and final state, $n$ and $n'$, respectively, refer to  spin-flip transitions between occupied and unoccupied states that include in addition a transition between the spin-orbit active states. Generalizing this thought means that for half-metallic chains, {\it i.e.}\ chains that have a band gap around the Fermi energy in one spin-channel,  a rather small DMI energy should be expected. It also explains the small DMI vectors  $D$ for the $3d$-Au chains recorded in Table~\ref{3d5d-MAE-DMI-t}. In case of Au chains the $d$-orbitals of Au that are responsible for the  spin-orbit matrix elements are all occupied and do not contribute to the DMI energy, while for Ir and Pt chains we have occupied and unoccupied states $5d$-orbitals, which make essential contributions to $D$.   

The strength of $\delta \varepsilon_n$ or $E_\textrm{DMI}$, respectively,  increases with increasing spin-canting angle $\varphi$ (through $\gamma$), increasing spin-orbit interaction $\xi$ and increasing hopping matrix elements $t_1$ and $t_2$, where in particular $t_2$ is proportional to the degree of structural asymmetry of the trimer.
Exactly this asymmetry and  both spin-mixing processes together are the factors responsible for the DMI. Due to the spin canting, the eigenstates are not anymore of pure spin character but contain a spin mixture of basis components contributing to non-zero spin-flip matrix elements at sites A, B, and to a superposition of $d_{xz}$ and $d_{yz}$ orbital character of different spin-character at site C and thus to the spin-orbit induced spin-flip on site C. Hence, the DMI is a non-local phenomenon since the state $\lvert n \rangle$ needs hybridization between the orbitals at the magnetic sites A and B and the non-magnetic site C carrying the spin-orbit interaction. Since $t/W$ is about $2t/|E_\mathrm{A}-\sigma\frac{1}{2} Im-E_\mathrm{C}|$ for small $t$ relative to the on-site energy differences and $\sqrt{2}\sgn{(t)}$ if $E_\mathrm{C}$ is in resonance with $E_\mathrm{A}-\sigma\frac{1}{2} Im$, we can conclude that the larger this hybridization either due to large hopping matrix elements $t$ or a  small energy difference between the on-site energies at the sites A and C the larger is $\delta \varepsilon_{n}$ explaining the strong dependence on the on-site energy difference in Fig.~\ref{compare_onsite_cases}. 

Regarding the sign of $\delta \varepsilon_{n}$, Eq.~\eqref{YM} gives some insight into the intricate relationship between the sign of the DMI and the underlying electronic structure even for this simple model. Apparently, the sign of the canting angle and the sign of the asymmetry of the trimer atoms through the hopping matrix element $t_2$ control directly the sign of the DMI energy. Further, the nature of the electronic structure in terms of the spin projection $\sigma$ of the occupied states, the sign of the hopping parameters $t_1$ and $t_2$ as well as the orbital character of the involved eigenstates $\lvert n^{\sigma} \rangle$  is crucial. In addition, the energetic position of $\varepsilon_n^{\sigma}$, $\varepsilon_{n'}^{\sigma'}$ and their bonding character have an influence on the sign, but also the magnitude of each term in the sum is important, making it not straightforward to relate  the sign of $\delta \varepsilon_{n}$ or $E_\textrm{DMI}$ to the physics of a system. Moreover, since the DMI energy is the integrated quantity over the values $\delta \varepsilon_{n}$ of all occupied states, the sign and the magnitude of the DMI depend on the magnitude and sign of all $\delta \varepsilon_{n}$.

To get a better understanding of the sign of the DMI on the basis of Eq.~\eqref{YM} we focus now on the sign of the energy shift $\delta \varepsilon$  in terms of  the DMI energy density $e_\mathrm{DM}$ and the DMI energy $E_\textrm{DMI}$. Namely, we will attempt to understand the behavior of $e_\mathrm{DM}$ and $E_\textrm{DMI}$
for $\varphi=45^\circ$ displayed in Fig.~\ref{noco_case}(b), in terms of the perturbation theory expression
Eq.~\eqref{YM} applied to unperturbed states for $\varphi=0^\circ$, shown in  Fig.~\ref{noco_case}(a).
 We concentrate first on the two low-lying pairs of occupied states $n$ in Fig.~\ref{noco_case}(a), the two majority states ($\sigma=1$) around $-1.75$~eV and the two minority states ($\sigma=-1$) around $-0.75$~eV. Both pairs are bonding states ($\tau=-1$) with $d_{xz}$ character of atoms A and B resulting from the hybridization with the orbitals at C. The lower (upper) peak of each pair results from the hybridization with $d_{yz}$, {\it i.e.}\ $i=1$, ($d_{xz}$,  $i=2$).  Since the states of each pair are of the same spin and exhibit a similar contribution of both orbitals at each site $d_{xz}^{\mathrm{A}}$ and $d_{xz}^{\mathrm{B}}$, the quantity $\delta S_{n^{\sigma'}n^\sigma}$ is approximately the same. The energy differences $\varepsilon_{n}^{\sigma}-\varepsilon_{n}^{\sigma'}$ for these two states to all other states $\lvert n^{\sigma'} \rangle$ are also almost the same, since each pair of states is well-separated from the other states. Hence, the only major difference turns out to be the sign in $\delta L_{n^\sigma n^{\sigma'}}$, which is determined by the orbital character of the states at the non-magnetic site and manifests as $(-1)^i$ in Eq.~\eqref{YM} and results at the end in a sign change of $e_\mathrm{DM}$ when passing through these peaks in energy.  

Now we focus on the pair of states $n$ and $n'$ around the Fermi energy (denoted as 1 and 2 in Fig.~\ref{noco_case}(a)), for which the energy dominator  $|\varepsilon_{n}^{\sigma} - \varepsilon_{n'}^{\sigma'}|$ is smallest and consequently whose contribution finally determines the DMI energy.  $n=(-1,1,\downarrow)$ is the highest occupied minority ($\sigma=-1$) state  with $d_{xz}$ ($i=1$) character and $n'=(-1,2,\uparrow)$ is lowest unoccupied majority state with $d_{yz}$ character ($i=2$). Both states are at the end of the eigenvalue spectrum and are therefore antibonding states ($\tau=\tau'=-1$). Recalling that $t_1>0$, $t_2<0$ and  $\varepsilon_{n}^{\sigma} <\varepsilon_{n'}^{\sigma'}$ we understand through~\eqref{YM} that $\delta \varepsilon$  is negative. To the DMI energy there contributes also a second pair of states of similar size but opposite in spin ($\sigma=1$), {\it i.e.}\ opposite sign,  $n=(-1,1,\uparrow)$ and $n=(-1,2,\downarrow)$, but their energy difference $|\varepsilon_{n}^{\sigma} - \varepsilon_{n'}^{\sigma'}|$ is slightly larger than the previous pair and thus the overall DMI energy is negative. Since the canting angle $\varphi=45^\circ$ produces a right-handed magnetic structure  with a chirality vector $\bm{c} = \bm{S_A} \times \bm{S_B} = - \bm{e_z}$, $D$ in this example is positive.

To conclude we developed a minimal model that carries general features of the DMI and is able to successfully reproduce and explain the DMI of the trimer regarding the symmetries, the magnitude and the sign. DMI can only occur at presence of spin-orbit interaction in inversion-asymmetric non-collinear magnetic systems and its driving force is the hybridization between the orbitals of the magnetic and non-magnetic sites.

\section{Conclusions}\label{sec-sum}

In the present paper, we have systematically investigated the non-collinear magnetic properties of infinite length 3$d$-5$d$ bi-atomic zigzag chains. Our investigations show that 3$d$-5$d$ chains exhibit an induced spin polarization on the 5$d$ atoms, which decreases with increasing atomic number of the 5$d$ element. In comparison to the Co-5$d$ chains, the magnetic moments of Fe-5$d$ chains show large variations as a function of lattice constant. We find a parabolic behavior of energy dispersion in the limit of large wave vectors $q$ for spin-spiral calculations without spin-orbit interaction. The ferromagnetic ($q=0$) and antiferromagnetic ($q=0.5$) calculations performed as special cases of the calculational model based on the spin-spiral concept were in good agreement with the conventional collinear ferromagnetic and antiferromagnetic calculations. Without inclusion of spin-orbit interaction, the Fe-Pt and Co-Pt chains exhibit a degenerate spin-spiral ground state at $q=\pm 0.07$ and $\pm 0.03$, respectively. 

Including the spin-orbit interaction, all 3$d$-5$d$ chains exhibit a non-vanishing DMI with signs that depend on the choice $5d$ metal, but only for the Fe-Pt and Co-Pt chains the DMI is sufficiently strong to compete with the MAE and the Heisenberg exchange to arrive at a non-collinear ground state. Since the non-collinear state is driven by the DMI, the magnetic structure is chiral in nature exhibiting a  homogeneously left-rotating cycloidal spin-spiral. The magnetic ground state of the Fe-Ir and Fe-Au chains remain unaffected by the DMI and exhibit a ferromagnetic ground state. 

We analyzed the behavior and strength of the  DMI on the basis of the electronic structure by means of the single particle energy. We observe strong shifts of the single particle energies due to the spin-orbit interaction in an energy regime, where $3d$ minority states hybridize with the $d$ states of the $5d$ metal.  Finite positive shifts of the energy eigenvalue followed by negative ones lead to positive and negative  contributions to the Dzyaloshinkii-Moriya interaction energy exhibiting an oscillating behavior of the DMI across the center of the 5$d$ band. Changing the $5d$ metal from Ir to Au moves the Fermi across the $5d$ band which explains the oscillatory sign of the DMI with the choice of the $5d$ metal. 

In order to provide a deeper understanding in the possible factors that influence the sign and strength of the DMI in low-dimensional systems on the level of the hybridization between relevant $d$-orbitals of the $3d$ and $5d$ atoms, we developed a minimal tight-binding model of a cluster of two magnetic $3d$ metal atoms and one non-magnetic $5d$ atom carrying the spin-orbit interaction assuming a triangular geometry. The model catches the main features of the {\it ab initio} results. The tight-binding calculations show that breaking of structural inversion symmetry and the non-collinearity of the magnetic sites are crucial to obtain a non-vanishing DMI. The strength of the DMI is linear in the strength of the spin-orbit interaction of atom $5d$. Further, the sign and strength  of DMI is also proportional the sign and strength of the hybridization between magnetic and non-magnetic sites and inversely proportional to the the energy difference between those states.

\begin{acknowledgments} 
We thank Gustav Bihlmayer and Frank Freimuth for fruitful discussions. These calculations were performed using $\textsf{\textit{JUROPA}}$ at J\"{u}lich Supercomputing Centre (JSC). V. K. acknowledges BARC-Mumbai, Forschungszentrum J\"{u}lich for financial assistance. We acknowledge DST and DAAD for travel funding support.
\end{acknowledgments}

\bibliography{dmi_kashid}

\begin{thebibliography}{49}
\expandafter\ifx\csname natexlab\endcsname\relax\def\natexlab#1{#1}\fi
\expandafter\ifx\csname bibnamefont\endcsname\relax
  \def\bibnamefont#1{#1}\fi
\expandafter\ifx\csname bibfnamefont\endcsname\relax
  \def\bibfnamefont#1{#1}\fi
\expandafter\ifx\csname citenamefont\endcsname\relax
  \def\citenamefont#1{#1}\fi
\expandafter\ifx\csname url\endcsname\relax
  \def\url#1{\texttt{#1}}\fi
\expandafter\ifx\csname urlprefix\endcsname\relax\def\urlprefix{URL }\fi
\providecommand{\bibinfo}[2]{#2}
\providecommand{\eprint}[2][]{\url{#2}}

\bibitem[{\citenamefont{Bode et~al.}(2007)\citenamefont{Bode, Heide, von
  Bergmann, Ferriani, Heinze, Bihlmayer, Kubetzka, Pietzsch, Bl\"{u}gel, and
  Wiesendanger}}]{Bode}
\bibinfo{author}{\bibfnamefont{M.}~\bibnamefont{Bode}},
  \bibinfo{author}{\bibfnamefont{M.}~\bibnamefont{Heide}},
  \bibinfo{author}{\bibfnamefont{K.}~\bibnamefont{von Bergmann}},
  \bibinfo{author}{\bibfnamefont{P.}~\bibnamefont{Ferriani}},
  \bibinfo{author}{\bibfnamefont{S.}~\bibnamefont{Heinze}},
  \bibinfo{author}{\bibfnamefont{G.}~\bibnamefont{Bihlmayer}},
  \bibinfo{author}{\bibfnamefont{A.}~\bibnamefont{Kubetzka}},
  \bibinfo{author}{\bibfnamefont{O.}~\bibnamefont{Pietzsch}},
  \bibinfo{author}{\bibfnamefont{S.}~\bibnamefont{Bl\"{u}gel}},
  \bibnamefont{and}
  \bibinfo{author}{\bibfnamefont{R.}~\bibnamefont{Wiesendanger}},
  \bibinfo{journal}{Nature} \textbf{\bibinfo{volume}{447}},
  \bibinfo{pages}{190} (\bibinfo{year}{2007}), ISSN \bibinfo{issn}{0028-0836}.

\bibitem[{\citenamefont{Heinze et~al.}(2000)\citenamefont{Heinze, Bode,
  Kubetzka, Pietzsch, Nie, Bl\"{u}gel, and Wiesendanger}}]{Heinze:00}
\bibinfo{author}{\bibfnamefont{S.}~\bibnamefont{Heinze}},
  \bibinfo{author}{\bibfnamefont{M.}~\bibnamefont{Bode}},
  \bibinfo{author}{\bibfnamefont{A.}~\bibnamefont{Kubetzka}},
  \bibinfo{author}{\bibfnamefont{O.}~\bibnamefont{Pietzsch}},
  \bibinfo{author}{\bibfnamefont{X.}~\bibnamefont{Nie}},
  \bibinfo{author}{\bibfnamefont{S.}~\bibnamefont{Bl\"{u}gel}},
  \bibnamefont{and}
  \bibinfo{author}{\bibfnamefont{R.}~\bibnamefont{Wiesendanger}},
  \bibinfo{journal}{Science} \textbf{\bibinfo{volume}{288}},
  \bibinfo{pages}{1805 } (\bibinfo{year}{2000}).

\bibitem[{\citenamefont{Ferriani et~al.}(2008)\citenamefont{Ferriani, von
  Bergmann, Vedmedenko, Heinze, Bode, Heide, Bihlmayer, Bl\"ugel, and
  Wiesendanger}}]{ferriani}
\bibinfo{author}{\bibfnamefont{P.}~\bibnamefont{Ferriani}},
  \bibinfo{author}{\bibfnamefont{K.}~\bibnamefont{von Bergmann}},
  \bibinfo{author}{\bibfnamefont{E.~Y.} \bibnamefont{Vedmedenko}},
  \bibinfo{author}{\bibfnamefont{S.}~\bibnamefont{Heinze}},
  \bibinfo{author}{\bibfnamefont{M.}~\bibnamefont{Bode}},
  \bibinfo{author}{\bibfnamefont{M.}~\bibnamefont{Heide}},
  \bibinfo{author}{\bibfnamefont{G.}~\bibnamefont{Bihlmayer}},
  \bibinfo{author}{\bibfnamefont{S.}~\bibnamefont{Bl\"ugel}}, \bibnamefont{and}
  \bibinfo{author}{\bibfnamefont{R.}~\bibnamefont{Wiesendanger}},
  \bibinfo{journal}{Phys. Rev. Lett.} \textbf{\bibinfo{volume}{101}},
  \bibinfo{pages}{027201} (\bibinfo{year}{2008}).

\bibitem[{\citenamefont{Menzel et~al.}(2012)\citenamefont{Menzel, Mokrousov,
  Wieser, Bickel, Vedmedenko, Bl\"ugel, Heinze, von Bergmann, Kubetzka, and
  Wiesendanger}}]{Menzel_12}
\bibinfo{author}{\bibfnamefont{M.}~\bibnamefont{Menzel}},
  \bibinfo{author}{\bibfnamefont{Y.}~\bibnamefont{Mokrousov}},
  \bibinfo{author}{\bibfnamefont{R.}~\bibnamefont{Wieser}},
  \bibinfo{author}{\bibfnamefont{J.~E.} \bibnamefont{Bickel}},
  \bibinfo{author}{\bibfnamefont{E.}~\bibnamefont{Vedmedenko}},
  \bibinfo{author}{\bibfnamefont{S.}~\bibnamefont{Bl\"ugel}},
  \bibinfo{author}{\bibfnamefont{S.}~\bibnamefont{Heinze}},
  \bibinfo{author}{\bibfnamefont{K.}~\bibnamefont{von Bergmann}},
  \bibinfo{author}{\bibfnamefont{A.}~\bibnamefont{Kubetzka}}, \bibnamefont{and}
  \bibinfo{author}{\bibfnamefont{R.}~\bibnamefont{Wiesendanger}},
  \bibinfo{journal}{Phys. Rev. Lett.} \textbf{\bibinfo{volume}{108}},
  \bibinfo{pages}{197204} (\bibinfo{year}{2012}).

\bibitem[{\citenamefont{Kubetzka et~al.}(2002)\citenamefont{Kubetzka, Bode,
  Pietzsch, and Wiesendanger}}]{Kubetzka_02}
\bibinfo{author}{\bibfnamefont{A.}~\bibnamefont{Kubetzka}},
  \bibinfo{author}{\bibfnamefont{M.}~\bibnamefont{Bode}},
  \bibinfo{author}{\bibfnamefont{O.}~\bibnamefont{Pietzsch}}, \bibnamefont{and}
  \bibinfo{author}{\bibfnamefont{R.}~\bibnamefont{Wiesendanger}},
  \bibinfo{journal}{Phys. Rev. Lett.} \textbf{\bibinfo{volume}{88}},
  \bibinfo{pages}{057201} (\bibinfo{year}{2002}),
  \urlprefix\url{http://link.aps.org/doi/10.1103/PhysRevLett.88.057201}.

\bibitem[{\citenamefont{Heide et~al.}(2008)\citenamefont{Heide, Bihlmayer, and
  Bl\"ugel}}]{heidem}
\bibinfo{author}{\bibfnamefont{M.}~\bibnamefont{Heide}},
  \bibinfo{author}{\bibfnamefont{G.}~\bibnamefont{Bihlmayer}},
  \bibnamefont{and} \bibinfo{author}{\bibfnamefont{S.}~\bibnamefont{Bl\"ugel}},
  \bibinfo{journal}{Phys. Rev. B} \textbf{\bibinfo{volume}{78}},
  \bibinfo{pages}{140403} (\bibinfo{year}{2008}).

\bibitem[{\citenamefont{Chen et~al.}(2013)\citenamefont{Chen, Zhu, Quesada, Li,
  N'Diaye, Huo, Ma, Chen, Kwon, Won et~al.}}]{YZWu_13}
\bibinfo{author}{\bibfnamefont{G.}~\bibnamefont{Chen}},
  \bibinfo{author}{\bibfnamefont{J.}~\bibnamefont{Zhu}},
  \bibinfo{author}{\bibfnamefont{A.}~\bibnamefont{Quesada}},
  \bibinfo{author}{\bibfnamefont{J.}~\bibnamefont{Li}},
  \bibinfo{author}{\bibfnamefont{A.~T.} \bibnamefont{N'Diaye}},
  \bibinfo{author}{\bibfnamefont{Y.}~\bibnamefont{Huo}},
  \bibinfo{author}{\bibfnamefont{T.~P.} \bibnamefont{Ma}},
  \bibinfo{author}{\bibfnamefont{Y.}~\bibnamefont{Chen}},
  \bibinfo{author}{\bibfnamefont{H.~Y.} \bibnamefont{Kwon}},
  \bibinfo{author}{\bibfnamefont{C.}~\bibnamefont{Won}}, \bibnamefont{et~al.},
  \bibinfo{journal}{Phys. Rev. Lett.} \textbf{\bibinfo{volume}{110}},
  \bibinfo{pages}{177204} (\bibinfo{year}{2013}).

\bibitem[{\citenamefont{Freimuth et~al.}(2014)\citenamefont{Freimuth, Bl\"ugel,
  and Mokrousov}}]{Freimuth_14}
\bibinfo{author}{\bibfnamefont{F.}~\bibnamefont{Freimuth}},
  \bibinfo{author}{\bibfnamefont{S.}~\bibnamefont{Bl\"ugel}}, \bibnamefont{and}
  \bibinfo{author}{\bibfnamefont{Y.}~\bibnamefont{Mokrousov}},
  \bibinfo{journal}{Journal of Physics: Condensed Matter}
  \textbf{\bibinfo{volume}{26}}, \bibinfo{pages}{104202}
  (\bibinfo{year}{2014}).

\bibitem[{\citenamefont{Ryu et~al.}(2013)\citenamefont{Ryu, Thomas, Yang, and
  Parkin}}]{Parkin_13}
\bibinfo{author}{\bibfnamefont{K.-S.} \bibnamefont{Ryu}},
  \bibinfo{author}{\bibfnamefont{L.}~\bibnamefont{Thomas}},
  \bibinfo{author}{\bibfnamefont{S.-H.} \bibnamefont{Yang}}, \bibnamefont{and}
  \bibinfo{author}{\bibfnamefont{S.}~\bibnamefont{Parkin}},
  \bibinfo{journal}{Nature Nanotechnology} \textbf{\bibinfo{volume}{8}},
  \bibinfo{pages}{527} (\bibinfo{year}{2013}).

\bibitem[{\citenamefont{Emori et~al.}(2013)\citenamefont{Emori, Bauer, Ahn,
  Martinez, and Beach}}]{Beach_13}
\bibinfo{author}{\bibfnamefont{S.}~\bibnamefont{Emori}},
  \bibinfo{author}{\bibfnamefont{U.}~\bibnamefont{Bauer}},
  \bibinfo{author}{\bibfnamefont{S.-M.} \bibnamefont{Ahn}},
  \bibinfo{author}{\bibfnamefont{E.}~\bibnamefont{Martinez}}, \bibnamefont{and}
  \bibinfo{author}{\bibfnamefont{G.~S.~D.} \bibnamefont{Beach}},
  \bibinfo{journal}{Nature Materials} \textbf{\bibinfo{volume}{12}},
  \bibinfo{pages}{611} (\bibinfo{year}{2013}).

\bibitem[{\citenamefont{Zakeri et~al.}(2010)\citenamefont{Zakeri, Zhang,
  Prokop, Chuang, Sakr, Tang, and Kirschner}}]{Kirschner_10}
\bibinfo{author}{\bibfnamefont{K.}~\bibnamefont{Zakeri}},
  \bibinfo{author}{\bibfnamefont{Y.}~\bibnamefont{Zhang}},
  \bibinfo{author}{\bibfnamefont{J.}~\bibnamefont{Prokop}},
  \bibinfo{author}{\bibfnamefont{T.-H.} \bibnamefont{Chuang}},
  \bibinfo{author}{\bibfnamefont{N.}~\bibnamefont{Sakr}},
  \bibinfo{author}{\bibfnamefont{W.~X.} \bibnamefont{Tang}}, \bibnamefont{and}
  \bibinfo{author}{\bibfnamefont{J.}~\bibnamefont{Kirschner}},
  \bibinfo{journal}{Phys. Rev. Lett.} \textbf{\bibinfo{volume}{104}},
  \bibinfo{pages}{137203} (\bibinfo{year}{2010}).

\bibitem[{\citenamefont{von Bergmann et~al.}(2006)\citenamefont{von Bergmann,
  Heinze, Bode, Vedmedenko, Bihlmayer, Bl\"ugel, and
  Wiesendanger}}]{Bergmann_06}
\bibinfo{author}{\bibfnamefont{K.}~\bibnamefont{von Bergmann}},
  \bibinfo{author}{\bibfnamefont{S.}~\bibnamefont{Heinze}},
  \bibinfo{author}{\bibfnamefont{M.}~\bibnamefont{Bode}},
  \bibinfo{author}{\bibfnamefont{E.~Y.} \bibnamefont{Vedmedenko}},
  \bibinfo{author}{\bibfnamefont{G.}~\bibnamefont{Bihlmayer}},
  \bibinfo{author}{\bibfnamefont{S.}~\bibnamefont{Bl\"ugel}}, \bibnamefont{and}
  \bibinfo{author}{\bibfnamefont{R.}~\bibnamefont{Wiesendanger}},
  \bibinfo{journal}{Phys. Rev. Lett.} \textbf{\bibinfo{volume}{96}},
  \bibinfo{pages}{167203} (\bibinfo{year}{2006}).

\bibitem[{\citenamefont{von Bergmann et~al.}(2007)\citenamefont{von Bergmann,
  Heinze, Bode, Bihlmayer, Bl{\"u}gel, and Wiesendanger}}]{Bergmann_07}
\bibinfo{author}{\bibfnamefont{K.}~\bibnamefont{von Bergmann}},
  \bibinfo{author}{\bibfnamefont{S.}~\bibnamefont{Heinze}},
  \bibinfo{author}{\bibfnamefont{M.}~\bibnamefont{Bode}},
  \bibinfo{author}{\bibfnamefont{G.}~\bibnamefont{Bihlmayer}},
  \bibinfo{author}{\bibfnamefont{S.}~\bibnamefont{Bl{\"u}gel}},
  \bibnamefont{and}
  \bibinfo{author}{\bibfnamefont{R.}~\bibnamefont{Wiesendanger}},
  \bibinfo{journal}{New Journal of Physics} \textbf{\bibinfo{volume}{9}},
  \bibinfo{pages}{396} (\bibinfo{year}{2007}).

\bibitem[{\citenamefont{Heinze et~al.}(2011)\citenamefont{Heinze, von Bergmann,
  Menzel, Brede, Kubetzka, Wiesendanger, Bihlmayer, and
  Bl{\"u}gel}}]{Heinze_11}
\bibinfo{author}{\bibfnamefont{S.}~\bibnamefont{Heinze}},
  \bibinfo{author}{\bibfnamefont{K.}~\bibnamefont{von Bergmann}},
  \bibinfo{author}{\bibfnamefont{M.}~\bibnamefont{Menzel}},
  \bibinfo{author}{\bibfnamefont{J.}~\bibnamefont{Brede}},
  \bibinfo{author}{\bibfnamefont{A.}~\bibnamefont{Kubetzka}},
  \bibinfo{author}{\bibfnamefont{R.}~\bibnamefont{Wiesendanger}},
  \bibinfo{author}{\bibfnamefont{G.}~\bibnamefont{Bihlmayer}},
  \bibnamefont{and}
  \bibinfo{author}{\bibfnamefont{S.}~\bibnamefont{Bl{\"u}gel}},
  \bibinfo{journal}{Nature Physics} \textbf{\bibinfo{volume}{7}},
  \bibinfo{pages}{713} (\bibinfo{year}{2011}).

\bibitem[{\citenamefont{Romming et~al.}(2013)\citenamefont{Romming, Hanneken,
  Menzel, Bickel, Wolter, von Bergmann, Kubetzka, and
  Wiesendanger}}]{Science_13}
\bibinfo{author}{\bibfnamefont{N.}~\bibnamefont{Romming}},
  \bibinfo{author}{\bibfnamefont{C.}~\bibnamefont{Hanneken}},
  \bibinfo{author}{\bibfnamefont{M.}~\bibnamefont{Menzel}},
  \bibinfo{author}{\bibfnamefont{J.~E.} \bibnamefont{Bickel}},
  \bibinfo{author}{\bibfnamefont{B.}~\bibnamefont{Wolter}},
  \bibinfo{author}{\bibfnamefont{K.}~\bibnamefont{von Bergmann}},
  \bibinfo{author}{\bibfnamefont{A.}~\bibnamefont{Kubetzka}}, \bibnamefont{and}
  \bibinfo{author}{\bibfnamefont{R.}~\bibnamefont{Wiesendanger}},
  \bibinfo{journal}{Science} \textbf{\bibinfo{volume}{341}},
  \bibinfo{pages}{636} (\bibinfo{year}{2013}).

\bibitem[{\citenamefont{Dzyaloshinskii}(1958)}]{Dzyaloshinsky:58.1}
\bibinfo{author}{\bibfnamefont{I.~E.} \bibnamefont{Dzyaloshinskii}},
  \bibinfo{journal}{Journal of Physics and Chemistry of Solids}
  \textbf{\bibinfo{volume}{4}}, \bibinfo{pages}{241 } (\bibinfo{year}{1958}),
  ISSN \bibinfo{issn}{0022-3697}.

\bibitem[{\citenamefont{Moriya}(1960)}]{moriya}
\bibinfo{author}{\bibfnamefont{T.}~\bibnamefont{Moriya}},
  \bibinfo{journal}{Phys. Rev.} \textbf{\bibinfo{volume}{120}},
  \bibinfo{pages}{91} (\bibinfo{year}{1960}).

\bibitem[{\citenamefont{Smith}(1976)}]{Smith:76.1}
\bibinfo{author}{\bibfnamefont{D.}~\bibnamefont{Smith}},
  \bibinfo{journal}{Journal of Magnetism and Magnetic Materials}
  \textbf{\bibinfo{volume}{1}}, \bibinfo{pages}{214 } (\bibinfo{year}{1976}),
  ISSN \bibinfo{issn}{0304-8853}.

\bibitem[{\citenamefont{Fert and Levy}(1980)}]{Fert:80.1}
\bibinfo{author}{\bibfnamefont{A.}~\bibnamefont{Fert}} \bibnamefont{and}
  \bibinfo{author}{\bibfnamefont{P.~M.} \bibnamefont{Levy}},
  \bibinfo{journal}{Phys. Rev. Lett.} \textbf{\bibinfo{volume}{44}},
  \bibinfo{pages}{1538} (\bibinfo{year}{1980}).

\bibitem[{\citenamefont{Fert}(1991)}]{Fert:90.1}
\bibinfo{author}{\bibfnamefont{A.}~\bibnamefont{Fert}}, in
  \emph{\bibinfo{booktitle}{Materials Science Forum}}
  (\bibinfo{organization}{Trans Tech Publ}, \bibinfo{year}{1991}),
  vol.~\bibinfo{volume}{59}, pp. \bibinfo{pages}{439--480}.

\bibitem[{\citenamefont{Kataoka et~al.}(1984)\citenamefont{Kataoka, Nakanishi,
  Yanase, and Kanamori}}]{Kataoka:84.1}
\bibinfo{author}{\bibfnamefont{M.}~\bibnamefont{Kataoka}},
  \bibinfo{author}{\bibfnamefont{O.}~\bibnamefont{Nakanishi}},
  \bibinfo{author}{\bibfnamefont{A.}~\bibnamefont{Yanase}}, \bibnamefont{and}
  \bibinfo{author}{\bibfnamefont{J.}~\bibnamefont{Kanamori}},
  \bibinfo{journal}{Journal of the Physical Society of Japan}
  \textbf{\bibinfo{volume}{53}}, \bibinfo{pages}{3624} (\bibinfo{year}{1984}),
  \urlprefix\url{http://jpsj.ipap.jp/link?JPSJ/53/3624/}.

\bibitem[{\citenamefont{Dzyaloshinskii}(1965)}]{Dzyaloshinsky:65:2}
\bibinfo{author}{\bibfnamefont{I.~E.} \bibnamefont{Dzyaloshinskii}},
  \bibinfo{journal}{Soviet Physics JETP} \textbf{\bibinfo{volume}{20}},
  \bibinfo{pages}{665} (\bibinfo{year}{1965}).

\bibitem[{\citenamefont{Tung and Wang}(2011)}]{Tung2011}
\bibinfo{author}{\bibfnamefont{J.}~\bibnamefont{Tung}} \bibnamefont{and}
  \bibinfo{author}{\bibfnamefont{Y.}~\bibnamefont{Wang}},
  \bibinfo{journal}{Journal of Magnetism and Magnetic Materials}
  \textbf{\bibinfo{volume}{323}}, \bibinfo{pages}{2032 }
  (\bibinfo{year}{2011}), ISSN \bibinfo{issn}{0304-8853}.

\bibitem[{\citenamefont{Che et~al.}(2012)\citenamefont{Che, Li, Ju, Chen, and
  Li}}]{che-2}
\bibinfo{author}{\bibfnamefont{X.~L.} \bibnamefont{Che}},
  \bibinfo{author}{\bibfnamefont{J.~H.} \bibnamefont{Li}},
  \bibinfo{author}{\bibfnamefont{H.~L.} \bibnamefont{Ju}},
  \bibinfo{author}{\bibfnamefont{X.~B.} \bibnamefont{Chen}}, \bibnamefont{and}
  \bibinfo{author}{\bibfnamefont{B.~H.} \bibnamefont{Li}},
  \bibinfo{journal}{Integrated Ferroelectrics} \textbf{\bibinfo{volume}{136}},
  \bibinfo{pages}{132} (\bibinfo{year}{2012}).

\bibitem[{\citenamefont{Negulyaev et~al.}(2013)\citenamefont{Negulyaev,
  Dorantes-D\'avila, Niebergall, Ju\'arez-Reyes, Pastor, and
  Stepanyuk}}]{Negulyaev_13}
\bibinfo{author}{\bibfnamefont{N.~N.} \bibnamefont{Negulyaev}},
  \bibinfo{author}{\bibfnamefont{J.}~\bibnamefont{Dorantes-D\'avila}},
  \bibinfo{author}{\bibfnamefont{L.}~\bibnamefont{Niebergall}},
  \bibinfo{author}{\bibfnamefont{L.}~\bibnamefont{Ju\'arez-Reyes}},
  \bibinfo{author}{\bibfnamefont{G.~M.} \bibnamefont{Pastor}},
  \bibnamefont{and} \bibinfo{author}{\bibfnamefont{V.~S.}
  \bibnamefont{Stepanyuk}}, \bibinfo{journal}{Phys. Rev. B}
  \textbf{\bibinfo{volume}{87}}, \bibinfo{pages}{054425}
  (\bibinfo{year}{2013}).

\bibitem[{\citenamefont{Wimmer et~al.}(1981)\citenamefont{Wimmer, Krakauer,
  Weinert, and Freeman}}]{FLAPW-1}
\bibinfo{author}{\bibfnamefont{E.}~\bibnamefont{Wimmer}},
  \bibinfo{author}{\bibfnamefont{H.}~\bibnamefont{Krakauer}},
  \bibinfo{author}{\bibfnamefont{M.}~\bibnamefont{Weinert}}, \bibnamefont{and}
  \bibinfo{author}{\bibfnamefont{A.~J.} \bibnamefont{Freeman}},
  \bibinfo{journal}{Phys. Rev. B} \textbf{\bibinfo{volume}{24}},
  \bibinfo{pages}{864} (\bibinfo{year}{1981}).

\bibitem[{\citenamefont{Jansen and Freeman}(1984)}]{FLAPW-2}
\bibinfo{author}{\bibfnamefont{H.~J.~F.} \bibnamefont{Jansen}}
  \bibnamefont{and} \bibinfo{author}{\bibfnamefont{A.~J.}
  \bibnamefont{Freeman}}, \bibinfo{journal}{Phys. Rev. B}
  \textbf{\bibinfo{volume}{30}}, \bibinfo{pages}{561} (\bibinfo{year}{1984}).

\bibitem[{fle()}]{fleur}
\bibinfo{note}{Http://www.flapw.de/}.

\bibitem[{\citenamefont{Herring et~al.}(1966)\citenamefont{Herring, Rado, and
  Suhl}}]{Herring:66.1}
\bibinfo{author}{\bibfnamefont{C.}~\bibnamefont{Herring}},
  \bibinfo{author}{\bibfnamefont{G.}~\bibnamefont{Rado}}, \bibnamefont{and}
  \bibinfo{author}{\bibfnamefont{H.}~\bibnamefont{Suhl}},
  \emph{\bibinfo{title}{Magnetism iv}} (\bibinfo{year}{1966}).

\bibitem[{\citenamefont{Sandratskii}(1986{\natexlab{a}})}]{Sandratskii:86.2}
\bibinfo{author}{\bibfnamefont{L.~M.} \bibnamefont{Sandratskii}},
  \bibinfo{journal}{physica status solidi (b)} \textbf{\bibinfo{volume}{136}},
  \bibinfo{pages}{167} (\bibinfo{year}{1986}{\natexlab{a}}), ISSN
  \bibinfo{issn}{1521-3951},
  \urlprefix\url{http://dx.doi.org/10.1002/pssb.2221360119}.

\bibitem[{\citenamefont{Sandratskii}(1991{\natexlab{a}})}]{Sandratskii:91.1}
\bibinfo{author}{\bibfnamefont{L.~M.} \bibnamefont{Sandratskii}},
  \bibinfo{journal}{Journal of Physics: Condensed Matter}
  \textbf{\bibinfo{volume}{3}}, \bibinfo{pages}{8565}
  (\bibinfo{year}{1991}{\natexlab{a}}),
  \urlprefix\url{http://stacks.iop.org/0953-8984/3/i=44/a=004}.

\bibitem[{\citenamefont{Zhang and Yang}(1998)}]{Zhang}
\bibinfo{author}{\bibfnamefont{Y.}~\bibnamefont{Zhang}} \bibnamefont{and}
  \bibinfo{author}{\bibfnamefont{W.}~\bibnamefont{Yang}},
  \bibinfo{journal}{Phys. Rev. Lett.} \textbf{\bibinfo{volume}{80}},
  \bibinfo{pages}{890} (\bibinfo{year}{1998}).

\bibitem[{\citenamefont{Vosko et~al.}(1980)\citenamefont{Vosko, Wilk, and
  Nusair}}]{Vosko}
\bibinfo{author}{\bibfnamefont{S.~H.} \bibnamefont{Vosko}},
  \bibinfo{author}{\bibfnamefont{L.}~\bibnamefont{Wilk}}, \bibnamefont{and}
  \bibinfo{author}{\bibfnamefont{M.}~\bibnamefont{Nusair}},
  \bibinfo{journal}{Canadian Journal of Physics} \textbf{\bibinfo{volume}{58}},
  \bibinfo{pages}{1200} (\bibinfo{year}{1980}).

\bibitem[{\citenamefont{Herring}(1966)}]{Herring}
\bibinfo{author}{\bibfnamefont{C.}~\bibnamefont{Herring}},
  \emph{\bibinfo{title}{Exchange interactions among itinerant electrons}}
  (\bibinfo{publisher}{Academic Press, New York, London},
  \bibinfo{year}{1966}).

\bibitem[{\citenamefont{Sandratskii}(1986{\natexlab{b}})}]{sandratskii}
\bibinfo{author}{\bibfnamefont{L.~M.} \bibnamefont{Sandratskii}},
  \bibinfo{journal}{physica status solidi (b)} \textbf{\bibinfo{volume}{136}},
  \bibinfo{pages}{167} (\bibinfo{year}{1986}{\natexlab{b}}), ISSN
  \bibinfo{issn}{1521-3951}.

\bibitem[{\citenamefont{Sandratskii}(1991{\natexlab{b}})}]{sandratskii2}
\bibinfo{author}{\bibfnamefont{L.~M.} \bibnamefont{Sandratskii}},
  \bibinfo{journal}{Journal of Physics: Condensed Matter}
  \textbf{\bibinfo{volume}{3}}, \bibinfo{pages}{8565}
  (\bibinfo{year}{1991}{\natexlab{b}}).

\bibitem[{\citenamefont{Heide et~al.}(2009)\citenamefont{Heide, Bihlmayer, and
  Bl\"{u}gel}}]{Heide}
\bibinfo{author}{\bibfnamefont{M.}~\bibnamefont{Heide}},
  \bibinfo{author}{\bibfnamefont{G.}~\bibnamefont{Bihlmayer}},
  \bibnamefont{and}
  \bibinfo{author}{\bibfnamefont{S.}~\bibnamefont{Bl\"{u}gel}},
  \bibinfo{journal}{Physica B: Condensed Matter}
  \textbf{\bibinfo{volume}{404}}, \bibinfo{pages}{2678 }
  (\bibinfo{year}{2009}), ISSN \bibinfo{issn}{0921-4526}.

\bibitem[{\citenamefont{Heide}(2006)}]{marcusthesis}
\bibinfo{author}{\bibfnamefont{M.}~\bibnamefont{Heide}}, Ph.D. thesis,
  \bibinfo{school}{RWTH Aachen University} (\bibinfo{year}{2006}).

\bibitem[{\citenamefont{Heide et~al.}(2011)\citenamefont{Heide, Bihlmayer, and
  Bl\"ugel}}]{Heide:2011:micmod}
\bibinfo{author}{\bibfnamefont{M.}~\bibnamefont{Heide}},
  \bibinfo{author}{\bibfnamefont{G.}~\bibnamefont{Bihlmayer}},
  \bibnamefont{and} \bibinfo{author}{\bibfnamefont{S.}~\bibnamefont{Bl\"ugel}},
  \bibinfo{journal}{Journal of Nanoscience and Nanotechnology}
  \textbf{\bibinfo{volume}{11}}, \bibinfo{pages}{3005} (\bibinfo{year}{2011}).

\bibitem[{\citenamefont{Kashid et~al.}(2011)\citenamefont{Kashid, Shah, and
  Salunke}}]{vikas1}
\bibinfo{author}{\bibfnamefont{V.}~\bibnamefont{Kashid}},
  \bibinfo{author}{\bibfnamefont{V.}~\bibnamefont{Shah}}, \bibnamefont{and}
  \bibinfo{author}{\bibfnamefont{H.}~\bibnamefont{Salunke}},
  \bibinfo{journal}{Journal of Nanoparticle Research}
  \textbf{\bibinfo{volume}{13}}, \bibinfo{pages}{5225} (\bibinfo{year}{2011}),
  ISSN \bibinfo{issn}{1388-0764}, \bibinfo{note}{10.1007/s11051-011-0507-8}.

\bibitem[{\citenamefont{Kashid et~al.}(2012)\citenamefont{Kashid, Shah, and
  Salunke}}]{vikas2}
\bibinfo{author}{\bibfnamefont{V.}~\bibnamefont{Kashid}},
  \bibinfo{author}{\bibfnamefont{V.}~\bibnamefont{Shah}}, \bibnamefont{and}
  \bibinfo{author}{\bibfnamefont{H.~G.} \bibnamefont{Salunke}},
  \bibinfo{journal}{AIP Conference Proceedings}
  \textbf{\bibinfo{volume}{1447}}, \bibinfo{pages}{465} (\bibinfo{year}{2012}).

\bibitem[{\citenamefont{Gay and Richter}(1986)}]{gay}
\bibinfo{author}{\bibfnamefont{J.~G.} \bibnamefont{Gay}} \bibnamefont{and}
  \bibinfo{author}{\bibfnamefont{R.}~\bibnamefont{Richter}},
  \bibinfo{journal}{Phys. Rev. Lett.} \textbf{\bibinfo{volume}{56}},
  \bibinfo{pages}{2728} (\bibinfo{year}{1986}).

\bibitem[{\citenamefont{Dorantes-D\'avila and Pastor}(1998)}]{dorantes}
\bibinfo{author}{\bibfnamefont{J.}~\bibnamefont{Dorantes-D\'avila}}
  \bibnamefont{and} \bibinfo{author}{\bibfnamefont{G.~M.}
  \bibnamefont{Pastor}}, \bibinfo{journal}{Phys. Rev. Lett.}
  \textbf{\bibinfo{volume}{81}}, \bibinfo{pages}{208} (\bibinfo{year}{1998}).

\bibitem[{\citenamefont{\'Ujfalussy et~al.}(1996)\citenamefont{\'Ujfalussy,
  Szunyogh, Bruno, and Weinberger}}]{Szunyo}
\bibinfo{author}{\bibfnamefont{B.}~\bibnamefont{\'Ujfalussy}},
  \bibinfo{author}{\bibfnamefont{L.}~\bibnamefont{Szunyogh}},
  \bibinfo{author}{\bibfnamefont{P.}~\bibnamefont{Bruno}}, \bibnamefont{and}
  \bibinfo{author}{\bibfnamefont{P.}~\bibnamefont{Weinberger}},
  \bibinfo{journal}{Phys. Rev. Lett.} \textbf{\bibinfo{volume}{77}},
  \bibinfo{pages}{1805} (\bibinfo{year}{1996}).

\bibitem[{\citenamefont{Slater and Koster}(1954)}]{SlaterKoster}
\bibinfo{author}{\bibfnamefont{J.~C.} \bibnamefont{Slater}} \bibnamefont{and}
  \bibinfo{author}{\bibfnamefont{G.~F.} \bibnamefont{Koster}},
  \bibinfo{journal}{Phys. Rev.} \textbf{\bibinfo{volume}{94}},
  \bibinfo{pages}{1498} (\bibinfo{year}{1954}).

\bibitem[{\citenamefont{Stoner}(1936)}]{Stoner_model1}
\bibinfo{author}{\bibfnamefont{E.~C.} \bibnamefont{Stoner}},
  \bibinfo{journal}{Proceedings of the Royal Society of London. Series A,
  Mathematical and Physical Sciences} \textbf{\bibinfo{volume}{154}},
  \bibinfo{pages}{pp. 656} (\bibinfo{year}{1936}), ISSN
  \bibinfo{issn}{00804630}.

\bibitem[{\citenamefont{Stoner}(1938)}]{Stoner_model2}
\bibinfo{author}{\bibfnamefont{E.~C.} \bibnamefont{Stoner}},
  \bibinfo{journal}{Proceedings of the Royal Society of London. Series A,
  Mathematical and Physical Sciences} \textbf{\bibinfo{volume}{165}},
  \bibinfo{pages}{pp. 372} (\bibinfo{year}{1938}), ISSN
  \bibinfo{issn}{00804630}.

\bibitem[{\citenamefont{Mukherjee and Cohen}(2001)}]{TB_noco}
\bibinfo{author}{\bibfnamefont{S.}~\bibnamefont{Mukherjee}} \bibnamefont{and}
  \bibinfo{author}{\bibfnamefont{R.}~\bibnamefont{Cohen}},
  \bibinfo{journal}{Journal of Computer-Aided Materials Design}
  \textbf{\bibinfo{volume}{8}}, \bibinfo{pages}{107} (\bibinfo{year}{2001}),
  ISSN \bibinfo{issn}{0928-1045}, \bibinfo{note}{10.1023/A:1020070028021}.

\bibitem[{\citenamefont{Aut\`{e}s et~al.}(2006)\citenamefont{Aut\`{e}s,
  Barreteau, Spanjaard, and Desjonquères}}]{TB_noco2}
\bibinfo{author}{\bibfnamefont{G.}~\bibnamefont{Aut\`{e}s}},
  \bibinfo{author}{\bibfnamefont{C.}~\bibnamefont{Barreteau}},
  \bibinfo{author}{\bibfnamefont{D.}~\bibnamefont{Spanjaard}},
  \bibnamefont{and} \bibinfo{author}{\bibfnamefont{M.-C.}
  \bibnamefont{Desjonquères}}, \bibinfo{journal}{Journal of Physics: Condensed
  Matter} \textbf{\bibinfo{volume}{18}}, \bibinfo{pages}{6785}
  (\bibinfo{year}{2006}).

\end{thebibliography}
\bibliographystyle{apsrev} 

\end{document}